\documentclass[11pt, twocolumn]{article}
\usepackage{geometry}
\usepackage{graphicx}
\usepackage{algorithm2e}
\RestyleAlgo{ruled}
\usepackage{caption}
\usepackage{subcaption}
\usepackage{amsmath}
\usepackage{url}
\usepackage{amssymb}
\usepackage[dvipsnames]{xcolor}
\usepackage{dsfont}
\usepackage{amssymb}
\usepackage{hyperref}
\hypersetup{colorlinks=true,
    linkcolor=black,
    filecolor=darkblue,      
    urlcolor=blue,
    citecolor=black }

\providecommand{\norm}[1]{\lVert#1\rVert}

\newcommand{\changes}[1]{\textcolor{black}{#1}}

\begin{document}
\title{\textbf{Observational cosmology with Artificial Neural Networks}}
\author{%
  Juan de Dios Rojas Olvera$^{\;1, a}$
  \and Isidro G\'omez-Vargas$^{\;2, b}$   
  \and J. Alberto V\'azquez$^{\;2, c, *}$
  }
  \date{%
    \small
    $^1$Facultad de Ciencias, Universidad Nacional Aut\'onoma de M\'exico, Ciudad de M\'exico, M\'exico\\%
    $^2$Instituto de Ciencias F\'isicas, Universidad Nacional Aut\'onoma de M\'exico, 62210, Cuernavaca, Morelos, M\'exico.\\[2ex]%
    \large
    \today \\[1ex]
    \small
    $^a$juan\_97dd\_@ciencias.unam.mx, $^b$igomez@icf.unam.mx, $^c$javazquez@icf.unam.mx$^*$
    }
    
\twocolumn[
  \begin{@twocolumnfalse}
\maketitle
    \begin{abstract}
    \changes{In cosmology, the analysis of observational evidence is very important to test theoretical models of the Universe. Artificial neural networks are powerful and versatile computational tools for data modelling and are recently being considered in the analysis of cosmological data. The main goal of this paper is to provide an introduction to artificial neural networks and to describe some applications to cosmology. We present an overview on the fundamentals of neural networks and their technical details. Throughout three examples, we show their capabilities in modelling cosmological data, saving computational time in numerical tasks, and classifying stellar objects. Artificial neural networks offer interesting qualities that make them a viable alternative method for data analysis in cosmological research. 
    }
 
    \end{abstract}
  \end{@twocolumnfalse}
]

\section{Introduction}

\changes{Observational cosmology has three relevant scientific pillars: cosmological theory, astronomical observations and statistical methods for data analysis. These areas allow us to compare theoretical cosmology with the observational evidence and thus validate or discard cosmological models.  
As the amount of observational data increases, the choice of data analysis methods becomes more crucial. Therefore, computational methods, including Machine Learning, have been incorporated into the cosmological field in recent years \cite{arjona2020can, wang2020machine, Chacon:2021sil}.}

\changes{Machine Learning is a branch of Artificial Intelligence focused on the mathematical modelling of the data. In recent times, the most growing field of Machine Learning is Deep Learning, which focuses on computational models called Artificial Neural Networks (ANNs). In several respects, neural networks have been proven to have more flexibility and computational power than other machine learning methods, so they have been successfully applied in a large number of fields ranging from industry and medicine to education and science, to name a few.}

\changes{There are several types of artificial neural networks and some of them fall under the category of supervised learning while others under the unsupervised learning. Hence, the problems that Deep Learning is able to solve include: regression, classification, pattern recognition, generative processes, time series, among many others. The ability of the ANNs to deal with complex and large datasets has allowed it to be a good alternative in several observational cosmological works \cite{lin2017does, peel2019distinguishing}, where theoretical models can be highly nonlinear, numerical methods be computationally expensive and there may be very complex datasets. }
\changes{Artificial neural networks have been used in several cosmological applications, such as N-body simulations \cite{rodriguez2018fast, he2019learning}, image analysis \cite{dieleman2015rotation, ntampaka2019deep}, statistical methods \cite{auld2007fast, alsing2019fast, li2019model}, and they have been also used to perform non-parametric reconstructions of cosmological functions \cite{dialektopoulos2021neural, gomez2021cosmological, wang2020reconstructing, escamilla2020deep}. In addition,} ANNs have enabled to decrease the computational time in cosmological calculations \cite{graff2012bambi, moss2020accelerated, hortua2020accelerating, gomez2021neural, mancini2021itcosmopower}.
Furthermore, the neural network implementation has made possible to analyse CMB signals \cite{baccigalupi2000neural, cmb_map}, and to classify observational measurements from extensive surveys, for example quasars in the Sloan Digital Sky Survey (SDSS) \cite{clasificacion_2018}. Finally, the use of deep learning in cosmology has increased considerably in recent years, and several works already incorporated this type of research, i.e. \cite{ribli2019improved, ishida2019machine, list2020galactic, dax2021real}.

The main goal of this paper is to present an introduction to deep learning followed by some examples of neural network applications to cosmology.
In Section \ref{sec:deeplearning} we present \changes{an overview of neural networks with the basic concepts}. Section \ref{sec:cosmology} contains the cosmological background necessary for the subsequent examples. In sections \ref{sec:ej_friedmann1}, \ref{sec:ej_ecdifs} and \ref{sec:ej_classification} we show three applications of the ANNs in cosmology. First, in Section \ref{sec:ej_friedmann1} we present a reconstruction of the Hubble parameter. Second, an ANN is trained from the solutions of the dynamical system of the Universe and its content (Section \ref{sec:ej_ecdifs}). Third, an example of stellar object classification is shown in Section \ref{sec:ej_classification}. Finally, in Section \ref{sec:conclusions} we provide some final comments and conclusions.

\section{\changes{A Deep Learning overview}}
\label{sec:deeplearning}

The initial steps of artificial neural networks started in 1943 when the first computational logic model for learning neurons was introduced \cite{mcculloch1943logical}. However, this model was only capable of solving a few linearly separable logic gates and demanding to provide plenty of information about the problem. Some years later, in 1957 Frank Rosenblatt \cite{rosenblatt1957perceptron} proposed the closest precursor to modern neural networks, the computational model known as the \textit{perceptron}.

The perceptron proposed changes in the learning process and unlike its predecessor, the neuron was now able to learn by itself from the input signals. However, in 1969 Marvin Minsky and Seymour Papert, famous Artificial Intelligence researchers, highlighted the inability of the perceptron to solve some non-linear problems, such as the well-known classification problem called \textit{exclusive OR} (XOR) \cite{minsky1969perceptron}.
Because of this, the scientific and technological community lost interest in neural networks and it took many years to change this situation. This stagnation stage is known as the \textit{artificial intelligence winter}. The curiosity for neural networks was regained in the 1980s, driven by Geoffrey Hinton and colleagues, who presented the \textit{backpropagation} algorithm that solved some of the computational limitations of neural networks \cite{rumelhart1986learning}. 
Nevertheless, it was until the 21st century, due to computing advancements, that neural networks regained great relevance, and even a new scientific discipline was born: \textit{Deep Learning}, focused exclusively on the Artificial Neural Networks study.

\changes{Nowadays there are several types of neural networks, for example, the Recurrent networks that are widely used in time series, the Convolutional neural nets that are very successful in image processing, the Autoencoders used in image denoising, and recently the Generative Adversarial networks. In this work, we focus on the most basic deep neural network which is known as the Multilayer Perceptron (MLP). }

\subsection{\changes{The Perceptron}}
\label{sec:perceptron}

The \textit{perceptron} is considered as the fundamental unit of neural networks, which consists of a mathematical model for a biological neuron. Neurons in the brain receive information from neighbour cells, process it and then, through synapses, transmit an electrical o chemical signal  to other neurons. The perceptron performs a similar task, it takes as an input $n$ signals, i.e. $x_1, x_2, \cdots, x_n$, each of them associated with a coefficient $w_j$, called \textit{weight}. Then, the perceptron computes a linear combination of weights and inputs, known as \textit{weighted sum} $z$:
\begin {equation}
  z = \sum_{j = 1}^n (x_jw_j) + b,
\end {equation}
where $b$ is called \textit{bias} and indicates the output value when all of the $x_i$ are null. The weights determine the strength of influence of each input on the neural network model.

Just like in biological neurons, a modulation of the signal $z$ is needed. In an ANN that modulation is carried out through the \textit{activation functions}, whose characteristics are described in \changes{Appendix} \ref{app:deeplearning_fnact}. The output of the perceptron is given by the activation function (denoted by $\sigma $) applied to the weighted sum:
\begin {equation}
    a = \sigma(z).
\end {equation}

\begin{figure}[t!]
\centering
\includegraphics[scale=0.35]{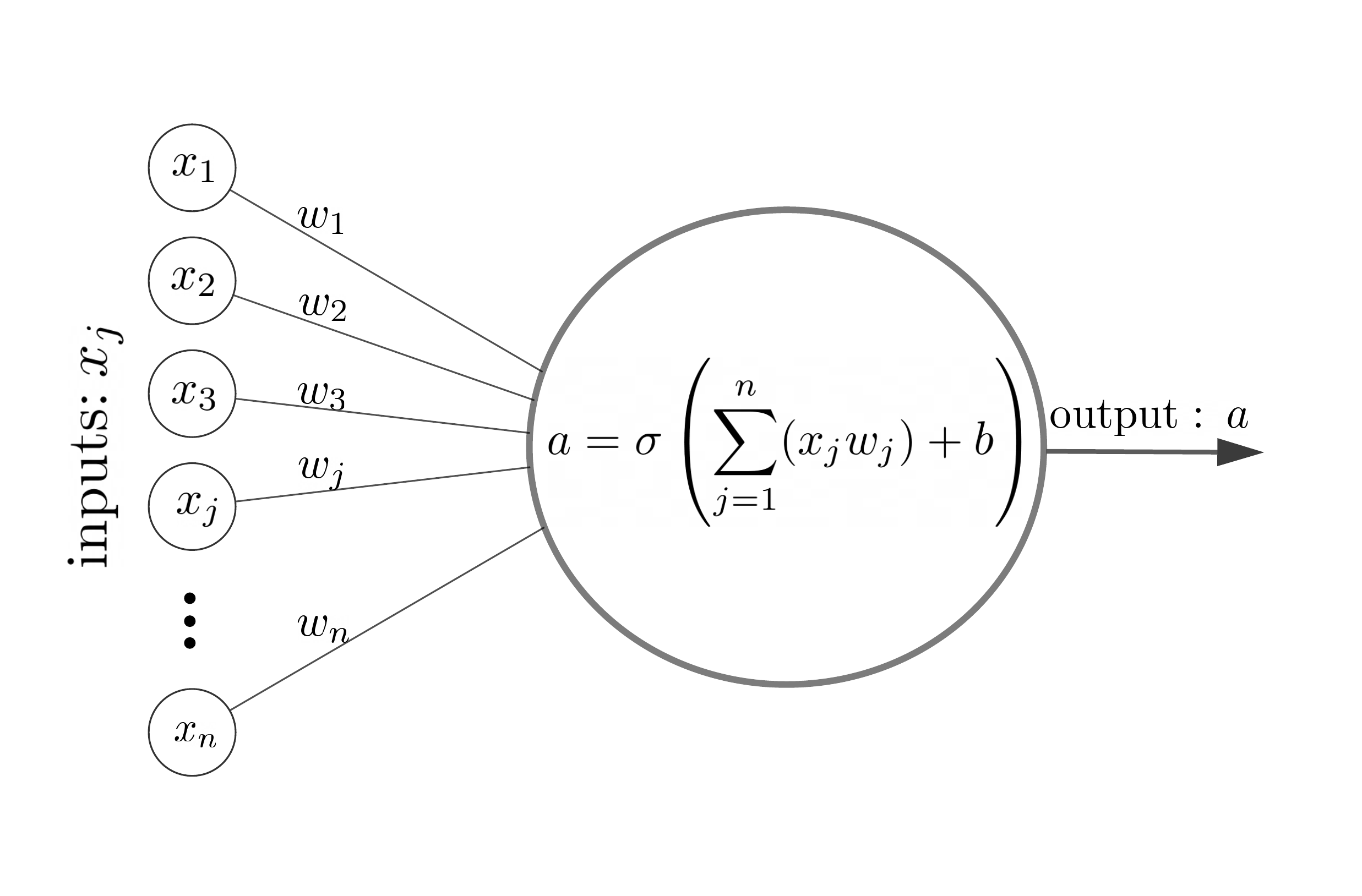}
\caption{\footnotesize{The perceptron processes the inputs and regularises them with an activation function, generating a numerical output.}}
\label{fig:percep}
\end{figure}

For a schematic description, see Fig. \ref{fig:percep}. This computational structure is commonly called a \textit{node} or \textit{neuron}, which acts as a brick for deep neural networks (multiple arrays of nodes). 
\\

\changes{The Algorithm \ref{alg:perceptron} shows the learning rule, for updating the weights, that the perceptron follows given a dataset made  of $x_i \in X$ independent variables and $y_i \in Y$ dependent variables. For the learning process the output of the perceptron should be compared with the target value included in the original dataset, this can be done with an error function called \textit{cost function}, in this case is the mean squared error between the target $y_i$ and the prediction $a_i$. 
The \textit{learning rate} $\eta$ indicates how much the weights change (see Appendix \ref{app:perceptron_descenso}). The iteration steps in the updating process are known as \textit{epochs}.}

\begin{algorithm}[t!]
\caption{The perceptron rule.}\label{alg:perceptron}

\KwData{$X=\{x_1, ..., x_n\}$, $Y=\{y_1, ..., y_n\}$.}

Generate random weights $w$.

\For{i in range{\rm (epochs)}}
{
    \For{$x_i \in X$ {\rm and} $y_i \in Y$}
    {
    \begin{itemize}
        \item Compute an output $a_i$.
    
        \item Update the weights:\\
        $w_i \longleftarrow w_i - \Delta w_i$, 
       
        where $\Delta w_i =\eta(a_i-y_i)\sigma'(z)x_i$,  
        \footnotesize{
        \begin{itemize}
            \item $\eta$: learning rate.
            \item $y_i$: true value of $f(x_i)$.
            \item $a_i$: perceptron prediction.
            \item $\sigma'$: activation function derivative.
        \end{itemize}}
    \end{itemize}
    }
}
\end{algorithm}

\subsection{\changes{Deep neural networks}}
\begin{figure}[t]
\centering
\includegraphics[scale=0.35]{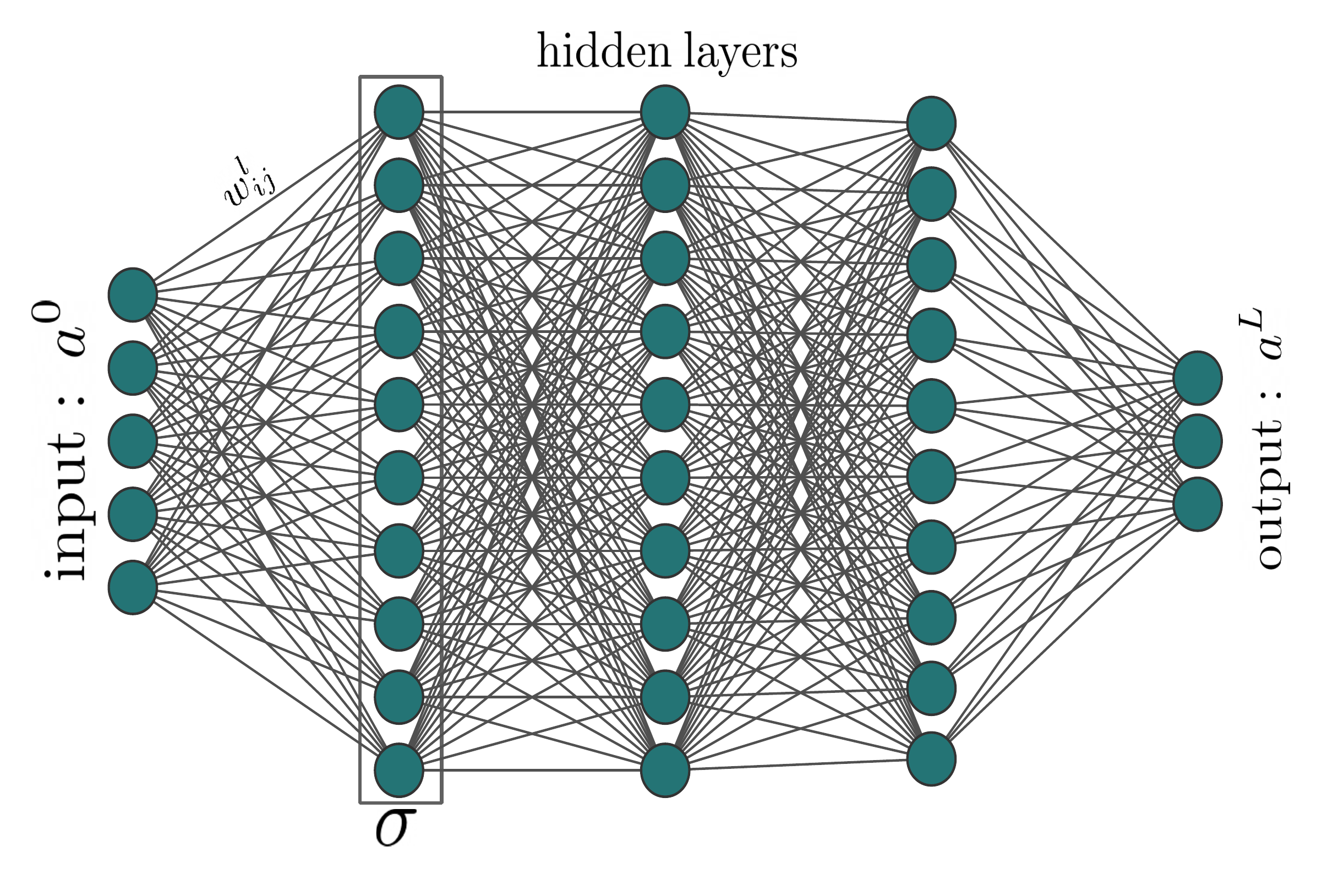}
\caption{\footnotesize{Structure of a deep neural network (or multilayer perceptron) with three hidden layers. Each neuron in the input layer receives a feature (or attribute) from the original dataset. The output layer generates a prediction that is compared, using a loss function, to the dependent variable within the original dataset.}}
\label{fig:red multicapa}
\end{figure}

The learning mechanism for a single neuron can be generalised to many of them and with a diverse type of arrangements.
The simplest deep neural network is known as the \textit{Multilayer Perceptron} (MLP) or \textit{feedforward neural network}, which consists of several perceptrons or nodes interconnected with each other through different arrangements of neurons called \textit{layers}. Every multilayer perceptron has an input layer; at least one intermediate layer, often called \textit {hidden layers}; and one output layer (see Fig. \ref{fig:red multicapa}). 
The input layer of an ANN corresponds to the data to be processed through the subsequent layers, and the number of nodes in this input layer must match the number of independent variables (features or attributes) of the dataset. Similarly to the perceptron, the output layer has to be compared with the dependent variables (labels or classes) of the dataset, which are the predicted ones.

The connections between one node to another have an associated weight $w_{ij}^l$, and a bias $b_{j}^l $, where $l$ represents the layer number. The weight connects the node $i$ in the layer $l-1$, with the node $j$ in the layer $l$. In consequence, the node $j$ in the layer $l$ will have an output $a_j^l$.
The weighted sum of each node contains information of the weights, biases and inputs of each layer, and is calculated as follows:
\begin{equation}
    z_j^l=\sum_i a_i^{l-1}w_{ij}^l + b^l_j.
\label{eq:sum ponderada entradas}
\end{equation}
Then, the activation functions $\sigma$ are applied to the weighted sum, Eq.~(\ref{eq:sum ponderada entradas}) and are denoted by:
\begin {equation}
     a_j^l = \sigma(z_j^l).
\label{eq:activacion entradas}
\end {equation}

Using the matrix notation allows us to establish the structure of the MLP in a simpler way.
Let $L$ be last layer, then the MLP output is the vector that takes each node $a_j^L$ as \changes{entries}:

\begin{equation}
    \begin{pmatrix}
    a^L_1\\
    a^L_2\\ 
    \vdots\\
    a_m^L
    \end{pmatrix}
    =a^L.
\label{eq:activation}
\end{equation}
Thus, we can call intermediate layers recursively as vectors $a^l$. For each layer, we can also build a weight matrix $W^l $ and denote the output $z^l$ of every layer as follows:
\begin{equation}
    z^l=a^{l-1}W^l+b^l.
\label{eq:activation capa anterior}
\end{equation}
Notice that the inputs of the associated weight matrix, with layer number $l$, are defined as: $$[W^l]_{ij}= w_{ij}^l,$$
with $b^l$ the vector that contains the biases of each node in the $l$ layer. Afterwards, it is necessary to apply the activation function $ \sigma$ as indicated in Eq.~(\ref{eq:activacion entradas}):
\begin{equation}
     a^l = \sigma(z^l).
\label{eq:sigma_vec}
\end{equation}
The process by which the information is transmitted, from the input layer of the neural network to the generation of an output, or prediction, in the last layer is known as \textit{forward propagation}. 
\\

A multilayer perceptron can be considered as a function $f:a^0 \in \mathds{R}^n \longrightarrow a^L \in \mathds{R}^k$. In this function, the hidden layers must also be considered even without being expressly mentioned.
The forward propagation of a MLP is represented in the following diagram:
\begin{equation}
\begin{split}
    a^0 \longrightarrow W^1a^0+b^1 \overset{\sigma}{\longrightarrow} a^1 \longrightarrow \cdots\\
\longrightarrow a^{l-1} \longrightarrow W^la^{l-1}+b^{l} \overset{\sigma}{\longrightarrow} a^L.
\end{split}
\end{equation}
For example, let us assume the structure of a neural network consists of a single hidden layer with $h$ nodes and a non-linear activation function $\sigma$. Using Eqs. (\ref{eq:activation capa anterior}) and (\ref{eq:sigma_vec}) we can calculate the hidden layer weighted sum $z^1$, as follows: 
$$z^1 = a^0W^1 +b^1.$$
\noindent
Here, the weight matrix $W^1$ is constructed by taking as a column $j$ the vector of weights that enter in the node $j$, thus in this case $W^1 \in \text{M}_{n\times h}$ and $b^1 \in \mathds{R}^h$. 
Now, we choose the activation function and apply it to each input of $z^1$:
$$\sigma(z^1)=a^1 \in \mathds{R}^h.$$
Finally we apply this process to the hidden layers, if any. In this case, for the output layer:
$$z^2=a^1W^2+b^2,$$
with $ W^2 \in \text{M}_{h \times k}$, $b^2 \in \mathds{R}^k$.
\\

Forward propagation is the first step in training a neural network. The output of forward propagation must be evaluated with an error function known as the cost function. Next, other mechanisms are necessary to update the neural network parameters and get the predictions closer to the target values. This will be discussed in the next section.

\subsection{Learning process}

\changes{Once the neural network has generated an output, it continues with the \textit{learning process}. The learning mechanism for a single neuron can be generalised to many of them \changes{and with a diverse type of arrangements}. Here the network parameters are updated so the output is as close as possible to the label values. The measure for this comparison will be indicated by the cost function; a usual cost function for regression tasks is the mean squared error (Eq. \ref{eq:error componentes}).}
\changes{Thus, the lower the value of the cost function, the better the neural network model. In deep learning, the most popular method for optimising cost functions is known as the \textit{gradient descent} (see Appendix \ref{app:perceptron_descenso} for details). When applying the gradient descent to update the network parameters in both the output layer and the hidden layers, an algorithm known as \textit{backpropagation} should be taken into account (Appendix \ref{app:backpropagation}). This algorithm indicates how to modify the parameters of the neural network such that the value of the cost function becomes smaller in an efficient way.}
\changes{The Algorithm \ref{alg:two} outlines the learning process of a MLP, which can be considered as a generalisation of Algorithm \ref{alg:perceptron}. The weights and bias updates are performed by applying the backpropagation Eqs. (\ref{eq:back 1}) - (\ref{eq:back 4}). A common way to initialise the random values for the weights and biases are random numbers normally distributed around zero.}

\begin{algorithm}[t]
\caption{Learning process.}\label{alg:two}
\KwData{$X$, $Y$.}
\textbf{Step 1}: generate random weights $W^1$ and biases $b^1$ for the network.

\For{i in range{\rm(epochs)}}
{

\textbf{Step 2}: compute an output of the ANN using forward propagation over $X$, then evaluate it with the cost function $C$ and the expected label $Y$.

\textbf{Step 3}: Update the network parameters with the backpropagation equations:
\begin{align*}
W^l  &\longrightarrow W^l-\eta \nabla_{W^l}C,\\
b^l  &\longrightarrow b^l-\eta \nabla_{b^l}C.    
\end{align*}
}
\end{algorithm}

\subsection{Overfitting and underfitting}
\label{sec:overfitting}

During the training process the neural networks tune the value of the weights to minimise the loss function. However, there may be several combinations of weights that similarly minimise the loss function. Therefore, a neural network must generate a good model for the data, capable of generalising and predicting values for the data not included in the training set. These machine learning issues have been rigorously studied in the branch of mathematics known as \textit{statistical learning theory}. In this work, we do not delve into these details, but for references see \cite{Ying_2019,avoid_overfit} and \cite[p.~142]{zhang2021dive}.
\\

In the ANN training, the dataset is usually split up into two parts: a \textit{training} set and a \textit{validation} set. The training set is used to train the neural network and therefore to fit the weights and biases. On the other hand, the validation set verifies whether the trained neural network managed to generalise beyond the dataset with it was trained. The split in the dataset allows to evaluate whether a neural network generates an acceptable model for the data. Then, we analyse the behaviour of the loss function and its change through the epochs by plotting the behaviour of the cost function on the training and validation sets. When a dataset is large enough, it is convenient to split it in three parts: the training set, the validation set and a test set; the last one is used to perform a final testing of the neural network performance.

\begin{figure}[t!]
     \centering
     \makebox[9cm][c]{
     \begin{subfigure}[t]{0.25\textwidth}
         \centering
         \includegraphics[width=\textwidth]{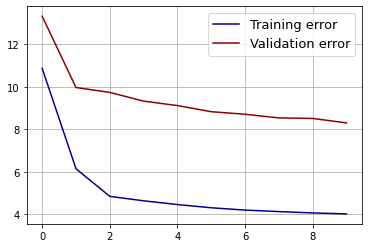}
         \caption{\scriptsize{The training error plateau and is not decreasing anymore, thus it keeps apart from the validation error.}}
         \label{fig:under_error}
     \end{subfigure}
     \begin{subfigure}[t]{0.25\textwidth}
         \centering
         \includegraphics[width=\textwidth]{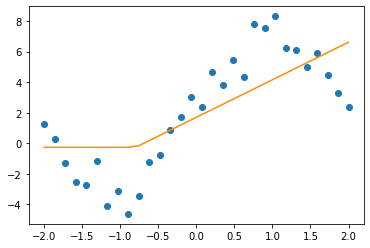}
         \caption{\scriptsize{The model (yellow line) is too simple, and hence not emulating the training set efficiently (blue dots).}}
         \label{fig:under_model}
     \end{subfigure}
     }
     \makebox[9cm][c]{
     \begin{subfigure}[t]{0.25\textwidth}
         \centering
         \includegraphics[width=\textwidth]{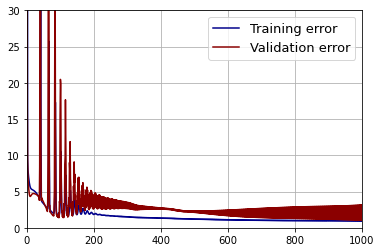}
         \caption{\scriptsize{The model was trained with too many epochs, and while the training error converges to zero, the validation error oscillates.}}
         \label{fig:over_error}
     \end{subfigure}
     \begin{subfigure}[t]{0.25\textwidth}
         \centering
         \includegraphics[width=\textwidth]{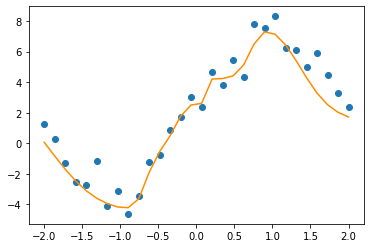}
         \caption{\scriptsize{The model was too well adapted to the training set that it is unable to generalise.}}
         \label{fig:over_model}
     \end{subfigure}
     }
     \makebox[9cm][c]{
     \begin{subfigure}[t]{0.25\textwidth}
         \centering
         \includegraphics[width=\textwidth]{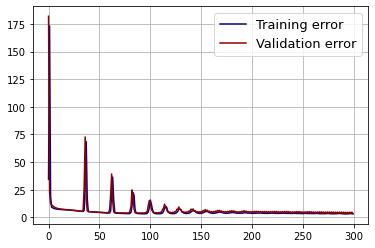}
         \caption{\scriptsize{The difference between the training curve and the validation is minimal, both converge to zero and the number of epochs is appropriate enough.}}
         \label{fig:optim_error}
     \end{subfigure}
     \begin{subfigure}[t]{0.25\textwidth}
         \centering
         \includegraphics[width=\textwidth]{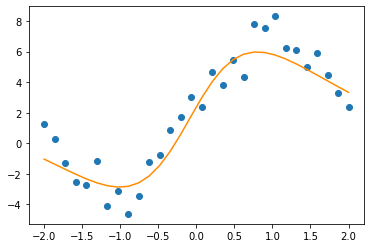}
         \caption{\scriptsize{A good model does not give up on training data, but emulates it well and can be generalised to more elements of the distribution.}}
     \label{fig:optim_model}
     \end{subfigure}
    } 
    \caption{\footnotesize{Comparison between different training processes and how we can infer information about model performance by plotting epochs ($x$-axis) with training and validation error.}}
    
    \label{fig:over_under}
\end{figure}

If throughout the epochs the model does not reduce the training error, this is a sign that the model is too simple, so it is not learning the pattern that underlies the training set. This phenomenon is called \textit{underfitting} (Fig. \ref{fig:under_error}) and it could be solved with a more complex model. On the other hand, if the training error is remarkably lower than the validation error, that is, the validation error does not decrease through the epochs while the training error does, this may be a sign that the model is unable to generalise what it has learned, this is called \textit{overfitting} (Fig. \ref{fig:over_error}). Overfitting can be solved with some processes called \textit{regularisation techniques}, some examples are dropout \cite{srivastava2014dropout} and regularisation norms \cite{louizos2017learning, phaisangittisagul2016analysis}.

\subsection{\changes{\textbf{Coding tips}}}
\label{mlp}

In this section \changes{we share some highlights about the neural network coding}. For technical details, \changes{we implemented a MLP from scracth, available} in \cite{git}.

The neural network architecture must be defined: the number of layers and the number of nodes. In addition, other intrinsic parameters (hyperparameters) of the neural network must also be established, such as the activation function, the number of epochs and the learning rate, among others. 
The hyperparameters must be calibrated so  the model generated by the neural network does not present problems such as underfitting or overfitting.
\changes{It is important that the cost function be small enough as required by the problem of interest. However, it is also necessary to take into account that the difference between the training and the validation errors should be small. By taking care of these aspects, the model generated by the neural network will be able to generalise to data not included in the original training data set and, therefore, make good predictions.} 

An interesting way to take advantage of a trained neural network model is to store it in a binary text file. In this way, the values of the weight matrices $W^l$ and the bias vectors $b^l$ obtained after proper training can be loaded to make more predictions, i.e., the neural network can be reusable. 

Building a neural network from scratch is a very good way to understand the fundamental concepts of Deep Learning. However, in practice it is necessary to work with different network architectures, several cost functions, activation functions and a large variety of hyperparameters; therefore it is best to use specialised Deep Learning libraries, for instance in \texttt{Python} such as \texttt{Pytorch}, \texttt{TensorFlow} or \texttt{Keras}, where the code is very efficient and contain a wide range of options. \changes{Other programming languages such as \texttt{R} and \texttt{Julia} also have their Deep Learning libraries.}

\section{\changes{Cosmological framework}} 
\label{sec:cosmology}

One of the most important equations in cosmology is the \textit{Friedmann equation}, which describes the evolution of the Universe and its expansion rate:
\begin{equation}
    H(t)^2  = \frac{\kappa_0}{3}\rho -\frac{kc^2}{a(t)^2},
    \label{eq:friedmann1}
\end{equation}
where $H$ is the Hubble parameter defined by $H(t) \equiv \dot{a}/a $ with
$a(t) $ the scale factor of the Universe; $c$ is the speed of light, $k$ is a constant that specifies the space geometry (curvature), and $ \kappa_0 $ is given in terms of the gravitational constant $G$, $ \kappa_0 = 8\pi G$.
Here $\rho$ represents the total energy density of the content of the cosmos, this is $\rho = \sum_i\rho_i$, where the index $i$ shows the possible components: radiation ($r$), baryonic and dark matter ($m$), and dark energy ($\Lambda$).

Another important equation in cosmology is the \textit{Continuity equation} or \textit{Fluid Equation}, 
which describes the behaviour and evolution of the content of the Universe
\begin{equation}
    \dot{\rho_i}+3\frac{\dot{a}}{a}\left(\rho_i + \frac{p_i}{c^2} \right)=0,
    \label{eq:fluido}
\end{equation}
where $p_i$ represents the pressure associated to every $i$ component.
A relationship between density and pressure can be established through an equation of state. The simplest way is to assume that the components behave as perfect fluids and they are described by a barotropic equation of state:
\begin{equation}
    p_i=(\gamma_i-1)\rho_i c^2, 
    \label{eq:estado}
\end{equation}
where $\gamma_i$ describes each fluid: radiation ($\gamma_r=4/3$), baryonic and dark matter ($\gamma_m = 1$) and dark energy in the form of cosmological constant ($ \gamma_\Lambda=0$). By substituting the value of $p_i$ in Eq.~(\ref{eq:fluido}), for each component, a system of couple differential equations is obtained
\begin{equation}
    \dot \rho_i + 3\gamma_iH\rho_i = 0.
\label{eq:diferenciales}
\end{equation}
Once we introduce the dimensionless \textit{density parameters}, defined as 
\begin{equation}
    \Omega_i = \frac{\kappa_0 }{3H^2} \rho_i,
    \label{eq:parametro_densidad}
\end{equation}
then, the set of Eqs.~(\ref{eq:diferenciales}) can be written as a dynamical system with the following form
\begin{equation}\label{eq:sys}
    \Omega_i' = 3(\Pi - \gamma_i)\Omega_i,
\end{equation}
with $\Pi = \sum_i \gamma_i \Omega_i$, and  prime notation means derivative with respect to the e-fold parameter $N = \ln(a) $. The Friedmann equation becomes a constraint for the density parameters at all time
\begin{equation} \label{eq:sum_omega}
    \sum_i \Omega_i= 1.
\end{equation}
This system can be solved traditionally, by setting some initial conditions for the parameters of radiation density, matter, dark energy and for the Hubble constant: $\Omega_{r,0}$, $\Omega_{m, 0}$, $\Omega_{\Lambda,0}$, $H_0$.

\section{\textbf{MLP applied to the Hubble parameter}}
\label{sec:ej_friedmann1}

\changes{In this example, we use a neural network to generate a model from 20 simulated data points \cite{gomez2021cosmological} corresponding to cosmic chronometers that consist of $H(z)$ measurements at redshift $z$ and their respective statistical error bars. This is a simple application that, in a more rigorous way, has been used in other research works \cite{escamilla2020deep, wang2020reconstructing, gomez2021cosmological, dialektopoulos2021neural}}.\\

\begin{figure}[t]
         \centering
         \includegraphics[trim = -30mm  0mm 50mm 0mm, clip, width=6.5cm, height=3.5cm]{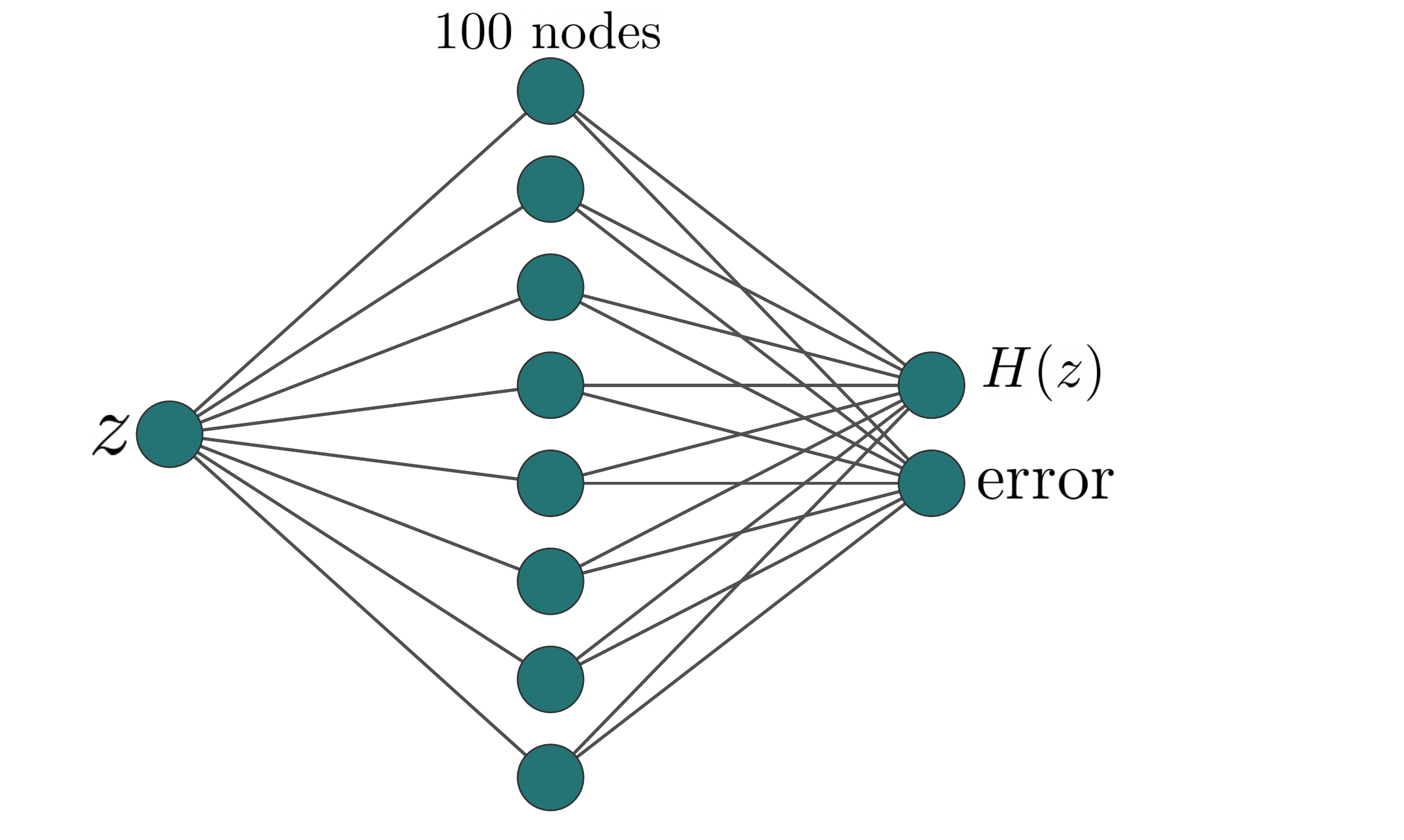}
         \includegraphics[trim = 3mm  0mm 0mm 0mm, clip, width=8.5cm, height=6.cm]{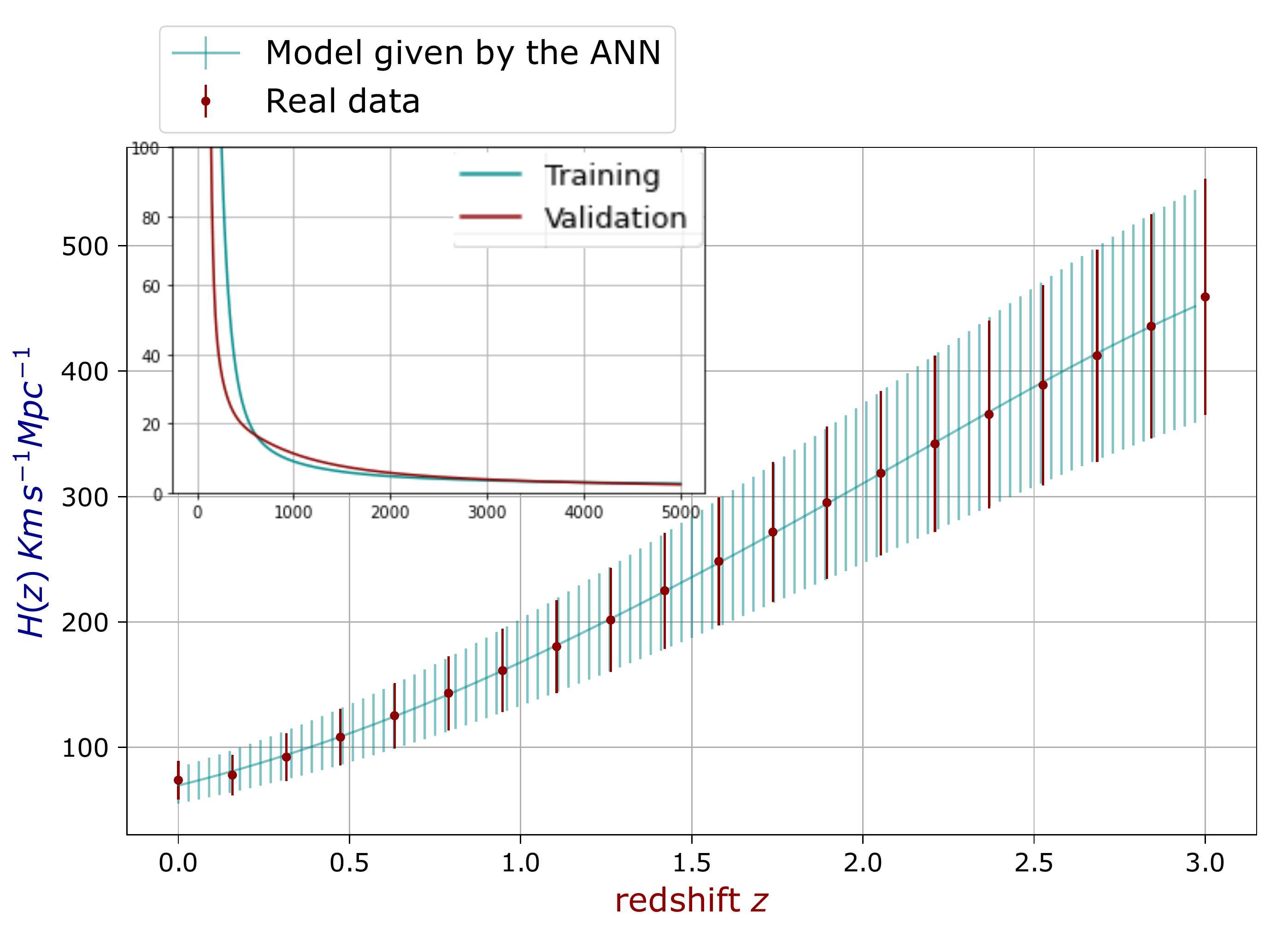}
         
         \caption{\footnotesize{Top: Neural network architecture. The input layer receives the redshift and the output layer generates a value for the Hubble parameter and its respective error. Bottom: Predictions of an ANN model for the Friedmann equation from a few data points. Red error bars are the original data used to train the architecture, while turquoise represent the predictions of the trained ANN. In the top left corner, the behaviour of the loss function in both the training and validation sets.}}
         \label{fig:fried+error}
\end{figure}

\changes{The neural network input corresponds to the redshift $z$ and the output to both $H$ and its error. Therefore, the neural network can be considered as a function from $z \in \mathds{R}$ $\to$ $\mathds{R}^2$. Top panel of Fig.~\ref{fig:fried+error} shows the architecture of the neural network.}

In this example, the trained neural network has one hidden layer with $100$ nodes, and it was calibrated with a total of $300$ trainable parameters (weights and bias). The dataset contains $20$ observational data points, however, the computational model generated by the neural network had a good reconstruction of the Hubble parameter without assuming any theoretical model beforehand; its cost function was $1.7$, this value is a bit large due to the small size of the data set, however it can be improved using more advanced strategies to choose the hyperparameters and to train the neural network \cite{gomez2021cosmological}.

\changes{The cost function curves of the training and validation sets show that their errors decreased at each epoch, which suggests that the model was able to learn from the data points provided and  generalise to any other missing $z$.} Once the training process was completed, the network is able to predict $H(z)$ values and their errors to values of redshift $z$ not included in the original dataset. See the lower panel of Fig.~\ref{fig:fried+error} for details.
\changes{In this figure, it can be noted that the neural network generated a model for the $H(z)$ measurements, and thus produced a reconstruction of the Hubble parameter based only on the dataset without making any cosmological assumptions beforehand. This result could be compared with the analytical $H(z)$ function coming from various cosmological models to assess how similar or different the curves are. That is, the MLP can be used to reconstruct various cosmological functions to obtain some physical conclusions \cite{Escamilla:2021uoj,Keeley:2020aym}.}

\section{\textbf{Cosmological differential equations}}
\label{sec:ej_ecdifs}

\changes{Solving a system of differential equations can be computationally demanding,
especially when this process has to be repeated multiple times.
The use of neural networks in solving systems of differential equations has already been addressed by different authors \cite{eqdiff,parciales1,parciales2, DUFERA2021100058}.}
\changes{In cosmology it is common that some cosmological functions have to be evaluated multiple times, for example in the case of simulations or in the Bayesian inference process \cite{graff2012bambi, mancini2021itcosmopower, hortua2020accelerating}. In this example, with various combinations of initial conditions, we solve the dynamical system to generate our training set. Then, we build the optimal architecture of a MLP to generate a model for this dataset.
Once the neural network is well trained, we can do the following with the predictions:}
\begin{itemize}
    \item \changes{Obtain the solutions of the system by evaluating the initial conditions at missing points in the training set.}
    \item \changes{Reduce the computational time to obtain multiple solutions.}
\end{itemize}
Throughout this example, we consider a flat Universe whose evolution goes back to the early times when radiation dominates. We also assume that its temperature is measured with great reliability, and this is reflected in a fix present radiation parameter  with value $\Omega_{r,0}=10^{-4}$.
Once this value is established, and since the sum of all densities must be equal to one (Eq. \ref{eq:sum_omega}), it is enough to vary the parameters $\Omega_{m,0}$ and $H_0$ to have the initial conditions for Eq.~(\ref{eq:sys}). With these parameters, the solution of the differential equations can be treated as a function $\Omega_i(N, \Omega_{m,0}, H_0)$.

To generate the set of solutions the following was performed: 
First, the intervals where the initial conditions are selected, correspond to  $N=\ln(a) \in [-12,0]$, $\Omega_{m,0} \in [0.1, 0.4]$ and $H_0 \in [65, 80]$. For the choice of these intervals, see \cite{Padilla:2019mgi}.
Then, the dataset $X$ was generated by computing the Cartesian product between all the intervals:  $X = [-12,0]\times [0.1,0.4] \times [65, 80]$, to get a set of about 30,000 elements.
Last, we compute the solutions corresponding to these elements $X$ to form the training dataset $Y$.

\begin{figure}[ht]
\centering
\includegraphics[trim = 0mm  0mm 0mm 0mm, clip, width=8.5cm, height=5.cm]{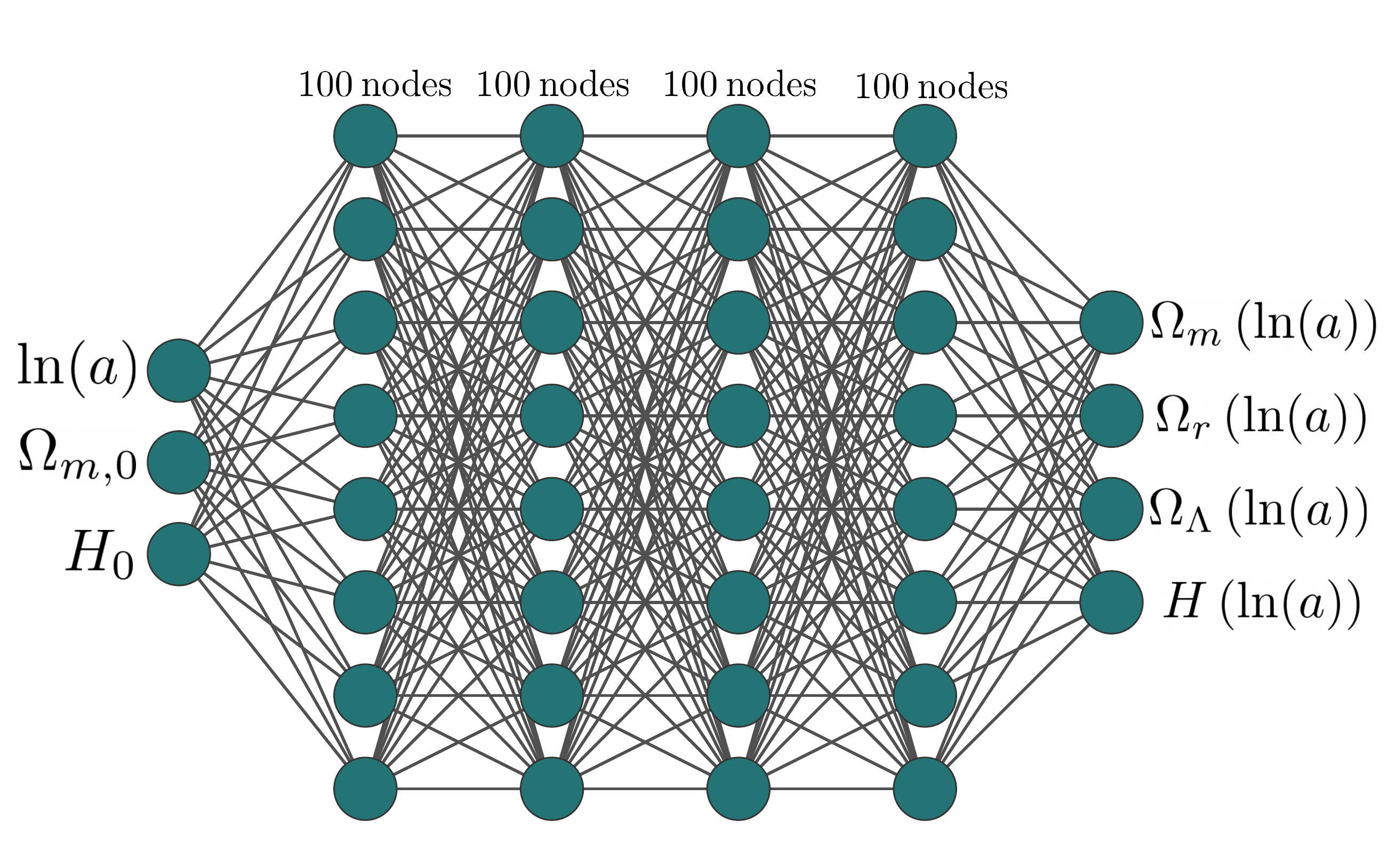}
\caption{\footnotesize{In this architecture the input corresponds to the initial conditions $\Omega_{m,0}$, $H_0$ and the domain variable $N=\ln(a)$, the MLP outputs are the solutions of each parameter $\Omega_i$ and $H$, in terms of $\ln(a)$.}}
\label{fig:arq3}
\end{figure}

\changes{The network architecture used to process these data can be seen in Fig.~\ref{fig:arq3}. Before starting the training process a max-min transformation was performed on the Y dataset, because the size of the Hubble factor becomes large for the early stages of the universe. 
The MLP architecture has four hidden layers with sigmoid activation function and considering both weights and bias they make a total of $69,154$ trainable parameters. It was trained for $500$ epochs with a dataset of $30,000$ elements and the final cost (error) function has a value of $2.9\times10^{-5}$.}

\begin{figure}[t]
\centering
\includegraphics[trim = 10mm  0mm 0mm 0mm, clip, width=9.2cm, height=5.3cm]{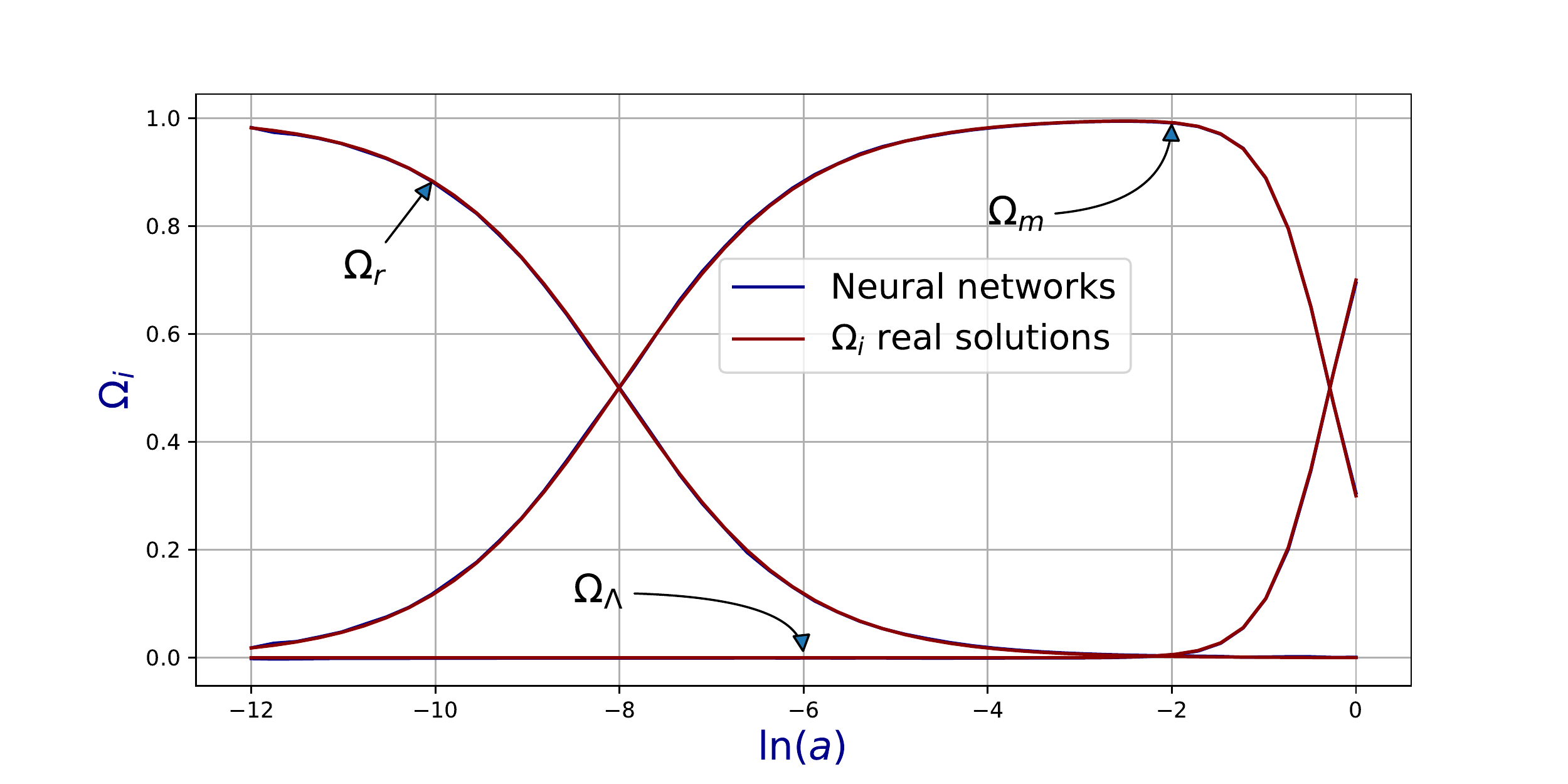}
\includegraphics[trim = 5mm  0mm 0mm 0mm, clip, width=8.5cm, height=5.3cm]{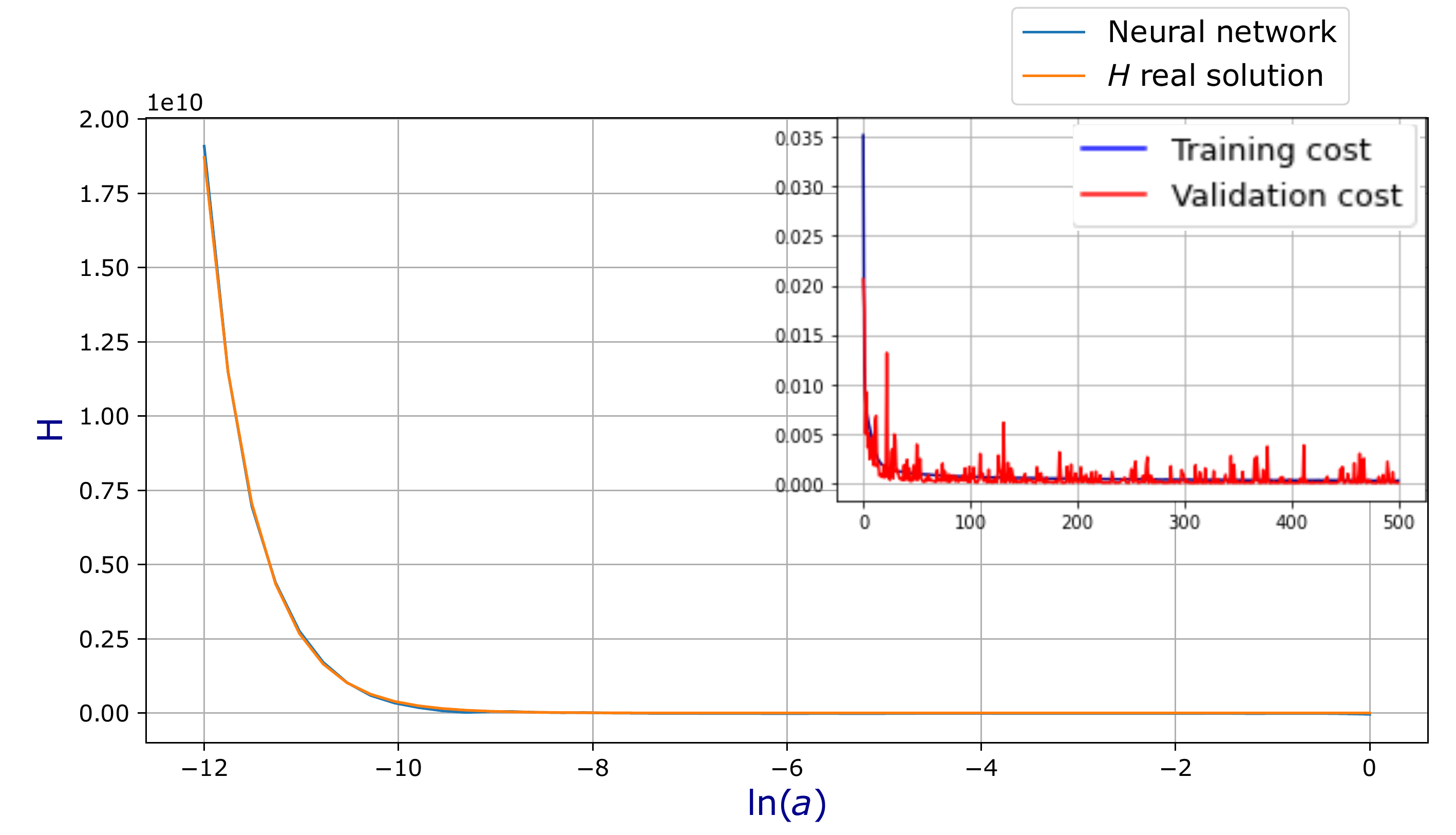}
\caption{\footnotesize{Comparison between the trained model and the real solutions, evaluated over the whole domain $\ln(a) \in [-12,0]$, with $\Omega_{m,0}=0.3$, $H_0=70$, 
for the density parameters (top) and the Hubble parameter (bottom).
In the upper right corner we can see the evolution of the error during the training process.}}
\label{fig:comparacion_sol}
\end{figure}

Considering a flat Universe components mentioned above, the dynamical system (\ref{eq:sys}) has analytical solutions, therefore the Friedmann equation, in terms of the redshift, $z=1/a-1$, takes the following form:
\begin{equation}
    \frac{H^2}{H_0^2} = \Omega_{r,0}(1+z)^4 + \Omega_{m,0}(1+z)^3 + \Omega_{\Lambda,0}.
    \label{eq:friedmann2}
\end{equation} 
Hence, in the top panel of Fig.~\ref{fig:comparacion_sol}, the solutions and the predictions of the neural network for the density parameters are compared, while in the bottom the results for the Hubble parameter $H$, \changes{ as well as the error curves, which show a good model performance. Notice the dynamical system in \ref{eq:sys} can be extended too a more complex that requires sophisticated numerical methods, i.e. \cite{Vazquez:2020ani, Gonzalez:2008wa}}

\begin{figure}[h!]
\centering
\includegraphics[trim = 15mm  0mm 0mm 0mm, clip, width=9.5cm, height=5.7cm]{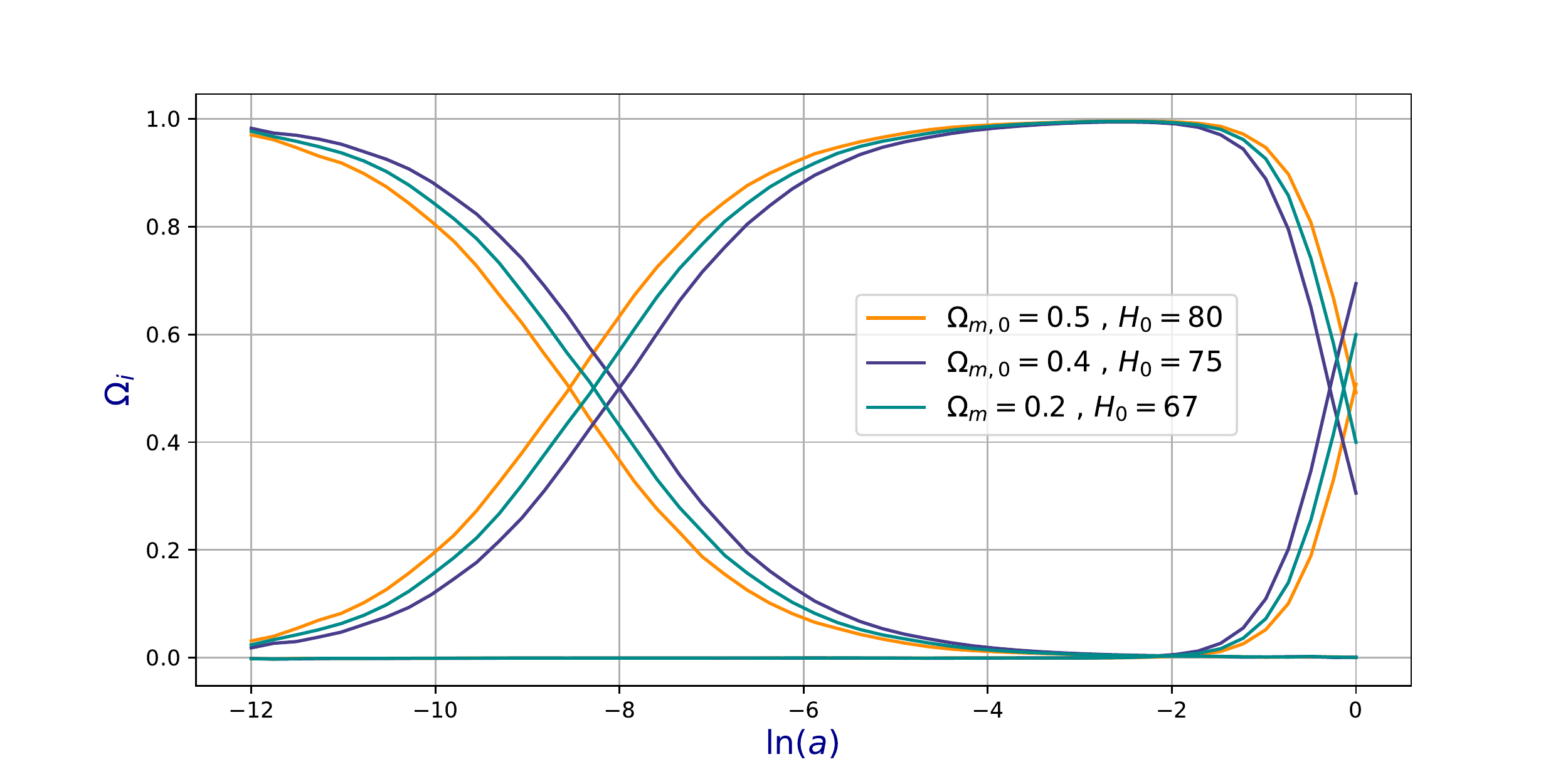}
\caption{\footnotesize{Predictions for different initial conditions, evaluating the ANN model instead of fully solving the differential equations.}}
\label{fig:varios_valores}
\end{figure}

Once the ANN was properly trained, we saved the generated model and used it to predict solutions of the system. \changes{But this time, it was enough to evaluate the model under a combination of initial conditions to obtain the solutions (Fig.~\ref{fig:varios_valores}).} Finally, the model was evaluated in $10,000$ different combinations of initial conditions and we found that the ANN reduces the computing time by $53\%$ compared to the numerical solutions of the differential equations. 

\section{\textbf{Classification of astronomical objects}}
\label{sec:ej_classification}
\changes{The classification of celestial objects is one of the main tasks that astronomers have carried out throughout history.  The increase in the flow of data we receive from the cosmos is both an opportunity and a challenge. It allows us to classify objects based on many of their characteristics, but the amount of information is so large that it can become a daunting task. This example shows that deep learning is a fantastic option to perform classification problems, in this case classification of stellar objects.}

In classification problems, the independent variables $X$ usually represent features or attributes, while the dependent variables $Y$ indicate the classes (or labels) that each item belongs to the dataset \cite{Chacon:2021sil}. Unlike the previous examples, the categorical (or classification) labels must be transformed to numeric elements called \textit{one-hot vectors}. It is convenient to apply the \textit{Softmax} activation function in the last layer, since this activation function makes it possible to associate the ANN output with the probability that the element being processed belongs to each of the existing classes in the dataset. 
The softmax function, applied to a vector with $k$ inputs $x_i$, is defined as:
\begin{equation}
\text{softmax}(x_i)=\frac{e^{x_i}}{\sum_{j=1}^k e^{x_j}}.
\label{eq:softmax}
\end{equation}
For this case we have used the \textit{Cross-Entropy} as a cost function, which is recommended for classification problems, and it is defined as follows:
\begin{equation}
C(Y,a^L)=-\sum_{j=1}^{k} Y_jln(a^L_j),
\label{eq:cross_entropy}
\end{equation}
therefore, this is the function to be minimised during neural network training. 

The dataset used for this example comes from the \textit{Sloan Digital Sky Survey DR14} \footnote{\url{https://www.sdss.org/dr14/}}, which consists of observations of different stellar objects: stars, quasars (QSO) and galaxies. These classes are represented in the $Y$ label set. On the other hand, there are 17 features which include from the redshift and the celestial coordinates, to characteristics of the spectrograph used; these attributes make up the features input $X$.

\begin{figure}[b]
\centering
\includegraphics[trim = 10mm  0mm 0mm 0mm, clip, width=7.9cm, height=5.cm]{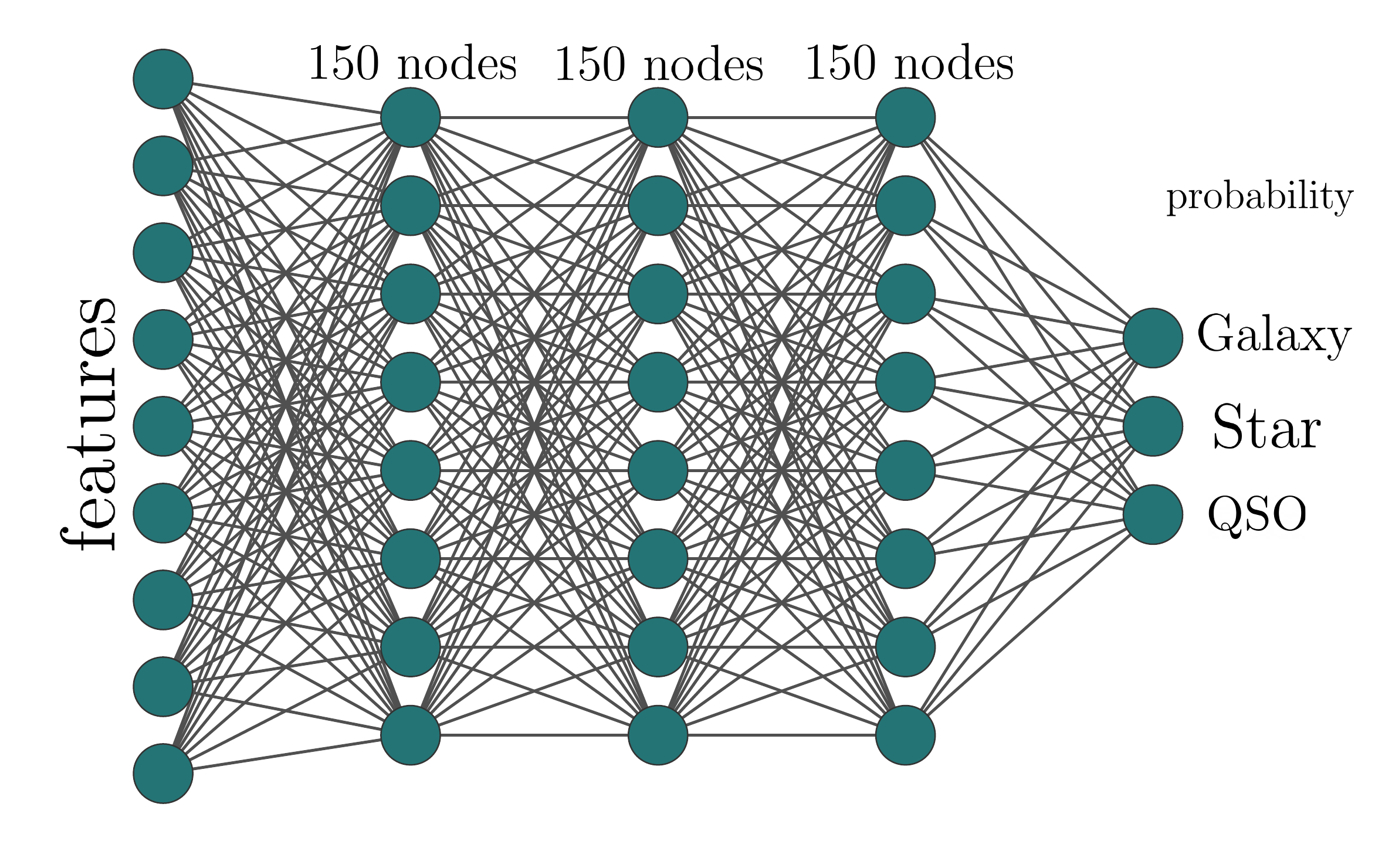}
\caption{\footnotesize{The ANN architecture used for the classification of astronomical objects. The input layer reads the attributes of the SDSS dataset and the output layer corresponds to the probability that each record is a given astronomical object, whether star, galaxy or quasar.}}
\label{fig:clasif}
\end{figure}

In Fig.~\ref{fig:clasif} we observe the ANN architecture used, with three hidden layers with sigmoid activation, and the output layer with the softmax activation function. Taking into account the weights and bias, this neural network has $70,803$ trainable parameters.

\begin{figure}[t]
\centering
\includegraphics[trim = 0mm  0mm 0mm 0mm, clip, width=7.9cm, height=5.cm]{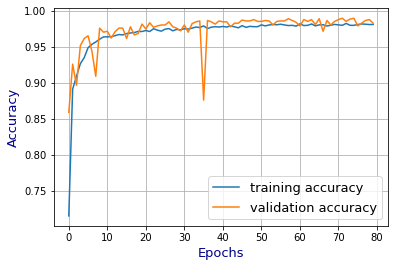}
\caption{\footnotesize{Behaviour of the accuracy metric over the epochs in the training and validation sets. It can be noticed that in both curves the accuracy is very similar and have a high value, therefore the ANN has been well trained.}}
\label{fig:acc_clasif}
\end{figure}

The ANN was trained over 80 epochs. To measure the performance in this classification task, we use the accuracy metric, defined as follows:
\begin{equation}
    \text{accuracy} = \frac{\#\text{Correct predictions}}{\#\text{ Total predictions}}.
\end{equation}
The ANN model, trained with the astronomical objects, reached 98.32\% accuracy on its training set and a 98.25\% for the validation set, as shown in Fig.~\ref{fig:acc_clasif}.

\begin{figure}[b]
\centering
\includegraphics[trim = 3mm  0mm 10mm 5mm, clip, width=8.5cm, height=6.cm]{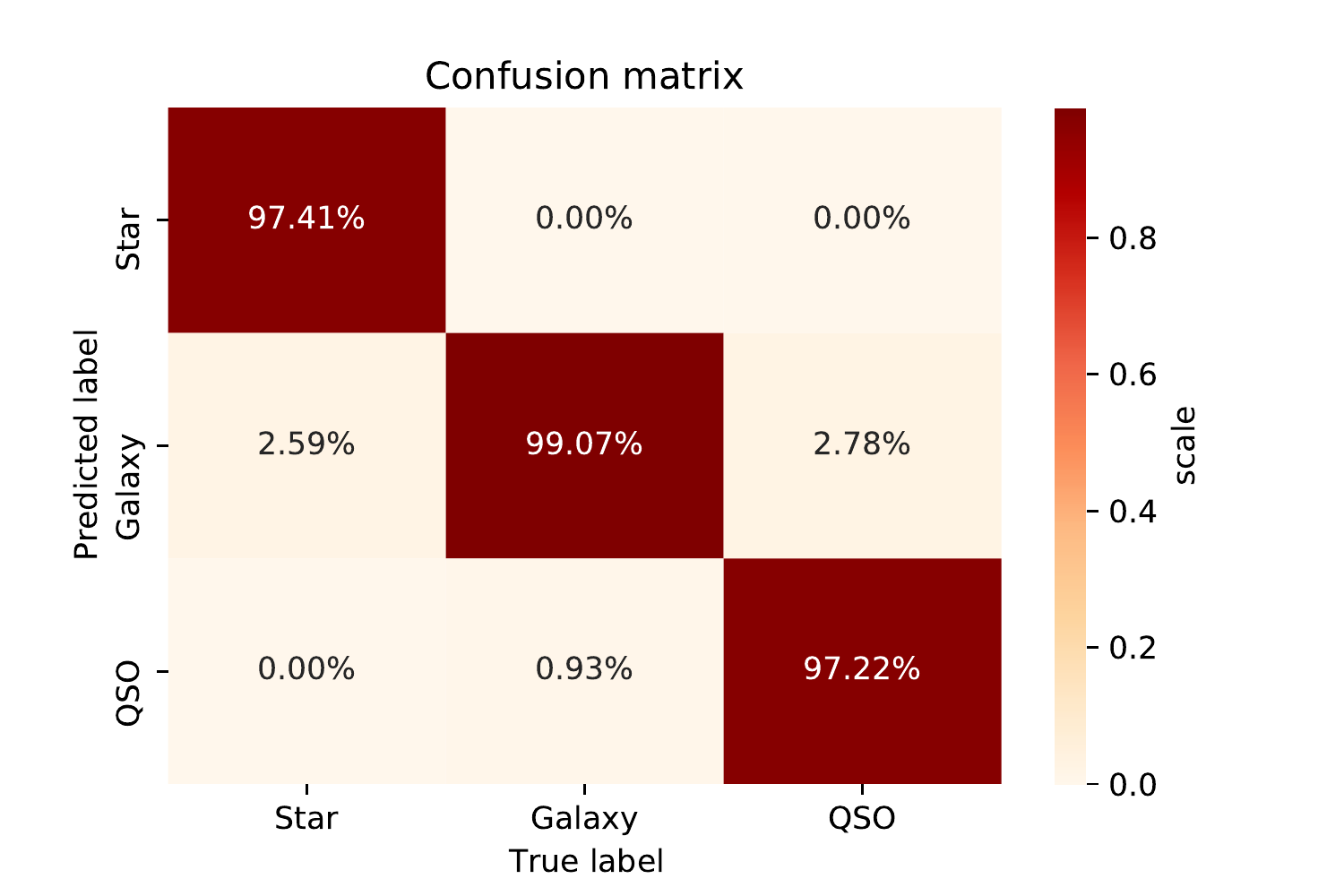}
\caption{\footnotesize{Confusion matrix of the classification model: percentage of accuracy for each of the classes in the test set.}}
\label{fig:matriz_conf}
\end{figure}

Finally, for testing the model efficacy, the ANN was asked to classify a set of $2,000$ elements outside of the training process, thus achieving a classification with 97.65 \% accuracy. Results of this classification can be organised in a confusion matrix (see Fig.~\ref{fig:matriz_conf}), which is an arrangement where the number of errors and successes of the ANN can be seen graphically. In a confusion matrix, the horizontal axis represents the real values and the vertical axis those predicted by the model,  such that the accuracy ratio is found on the diagonal matrix, and the mistakes ratio out of it. It can be noticed that, in this example, the errors are minimal compared to the correct data labelling done by the ANN.

\section{Conclusions}
\label{sec:conclusions}
This paper presented an introduction to the fundamental concepts of Deep Learning, \changes{with the intention of providing new tools for cosmological analysis.} The applications were presented through three practical examples,
and through them we showed that:

\begin{itemize}
\item Neural networks have the ability to emulate any function or pattern that fulfils a given data set. This can be very useful in various scientific areas, because the networks can generate a computational model for the data and are a good alternative when a satisfactory analytical model is not available.

\item A properly trained neural network can be used to replace traditional computational calculations in a wide variety of problems and thus decrease the computational time. In addition, in section \ref{sec:ej_ecdifs} we showed that the neural network also provides a model that can be evaluated and mathematically manipulated, something that traditional numerical methods may not always offer.

\item With the last example, we showed the great efficiency of neural networks in tasks that can be complicated to perform, for example objects classification. In our case, we performed a classification with numerical features, however, the literature indicates that neural networks are also a great tool in image and video classification.
\end{itemize}

The examples that we presented in this article are just a small sample of the potential that Deep Learning can have in cosmology. It is a branch of Artificial Intelligence that is booming and that, little by little, is being incorporated into various scientific disciplines.

\changes{Even though the ANNs are considered as a great tool for several problems, in many cases they may present some  disadvantages. For example: the training time can be considerably long for a very large database. It is commonly said that neural networks are 'black boxes', because the huge number of parameters that are just a set of real numbers with no information about the phenomenon they are modelling. The ANN must be used with caution to prevent  overfitting or underfitting to ensure that the model can be generalised.}
\changes{Continuing with this line of work, more deep learning tools can also be presented to facilitate data management in science, mainly in astronomy and observational cosmology. A next step can be found in numerical simulations of the universe and its contents or convolutional neural networks, widely used in image processing. However, that's a topic for another time.}

\subsection*{Acknowledgments}
 J.A.V. acknowledges the support provided by FOSEC SEP-CONACYT Investigaci\'on B\'asica A1-S-21925, PRONACES-CONACYT/304001/2020, and UNAM-DGAPA-PAPIIT IA104221. IGV thanks the CONACYT postdoctoral grant and ICF-UNAM.\\
 JDD.R.O Thanks to ICF-UNAM, and UNAM-DGAPA-PAPIIT IA102219 and IA104221, for their support.

\bibliographystyle{JHEP}
\bibliography{bibliography.bib}

\providecommand{\href}[2]{#2}\begingroup\raggedright\begin{thebibliography}{10}

\bibitem{arjona2020can}
R.~Arjona and S.~Nesseris, \emph{What can machine learning tell us about the
  background expansion of the universe?},
  \href{https://doi.org/https://doi.org/10.1103/PhysRevD.101.123525}{\emph{Physical
  Review D} {\bfseries 101} (2020) 123525}.

\bibitem{wang2020machine}
G.-J.~Wang, X.-J.~Ma and J.-Q.~Xia, \emph{Machine learning the cosmic curvature
  in a model-independent way},
  \href{https://doi.org/https://doi.org/10.1093/mnras/staa4044}{\emph{Monthly
  Notices of the Royal Astronomical Society} {\bfseries 501} (2021) 5714}.

\bibitem{Chacon:2021sil}
J.~Chac\'on, J.A.~V\'azquez and E.~Almaraz, \emph{{Classification algorithms
  applied to structure formation simulations}},
  \href{https://arxiv.org/abs/2106.06587}{{\ttfamily 2106.06587}}.

\bibitem{lin2017does}
H.W.~Lin, M.~Tegmark and D.~Rolnick, \emph{Why does deep and cheap learning
  work so well?},
  \href{https://doi.org/https://doi.org/10.1007/s10955-017-1836-5}{\emph{J.
  Statistical Physics} {\bfseries 168} (2017) 1223}.

\bibitem{peel2019distinguishing}
A.~Peel, F.~Lalande, J.-L.~Starck, V.~Pettorino, J.~Merten, C.~Giocoli et~al.,
  \emph{Distinguishing standard and modified gravity cosmologies with machine
  learning},
  \href{https://doi.org/https://doi.org/10.1103/PhysRevD.100.023508}{\emph{Physical
  Review D} {\bfseries 100} (2019) 023508}.

\bibitem{rodriguez2018fast}
A.C.~Rodr{\'\i}guez, T.~Kacprzak, A.~Lucchi, A.~Amara, R.~Sgier, J.~Fluri
  et~al., \emph{Fast cosmic web simulations with generative adversarial
  networks},
  \href{https://doi.org/https://doi.org/10.1186/s40668-018-0026-4}{\emph{Comp.
  Astrophys. and Cosmology} {\bfseries 5} (2018) 4}.

\bibitem{he2019learning}
S.~He, Y.~Li, Y.~Feng, S.~Ho, S.~Ravanbakhsh, W.~Chen et~al., \emph{Learning to
  predict the cosmological structure formation},
  \href{https://doi.org/https://doi.org/10.1073/pnas.1821458116}{\emph{Proc.
  Natl. Acad. Sci.} {\bfseries 116} (2019) 13825}.

\bibitem{dieleman2015rotation}
S.~Dieleman, K.W.~Willett and J.~Dambre, \emph{Rotation-invariant convolutional
  neural networks for galaxy morphology prediction},
  \href{https://doi.org/https://doi.org/10.1093/mnras/stv632}{\emph{Monthly
  Notices of the Royal Astronomical Society} {\bfseries 450} (2015) 1441}.

\bibitem{ntampaka2019deep}
M.~Ntampaka, J.~ZuHone, D.~Eisenstein, D.~Nagai, A.~Vikhlinin, L.~Hernquist
  et~al., \emph{A deep learning approach to galaxy cluster x-ray masses},
  \href{https://doi.org/https://doi.org/10.3847/1538-4357/ab14eb}{\emph{The
  Astrophysical Journal} {\bfseries 876} (2019) 82}.

\bibitem{auld2007fast}
T.~Auld, M.~Bridges, M.~Hobson and S.~Gull, \emph{Fast cosmological parameter
  estimation using neural networks},
  \href{https://doi.org/https://doi.org/10.1111/j.1745-3933.2006.00276.x}{\emph{Monthly
  Notices of the Royal Astronomical Society: Letters} {\bfseries 376} (2007)
  L11}.

\bibitem{alsing2019fast}
J.~Alsing, T.~Charnock, S.~Feeney and B.~Wandelt, \emph{Fast likelihood-free
  cosmology with neural density estimators and active learning},
  \href{https://doi.org/https://doi.org/10.1093/mnras/stz1960}{\emph{Monthly
  Notices of the Royal Astronomical Society} {\bfseries 488} (2019) 4440}.

\bibitem{li2019model}
S.-Y.~Li, Y.-L.~Li and T.-J.~Zhang, \emph{Model comparison of dark energy
  models using deep network},
  \href{https://doi.org/https://doi.org/10.1088/1674-4527/19/9/137}{\emph{Res.
  Astron. Astrophys.} {\bfseries 19} (2019) 137}.

\bibitem{dialektopoulos2021neural}
K.~Dialektopoulos, J.L.~Said, J.~Mifsud, J.~Sultana and K.Z.~Adami,
  \emph{Neural network reconstruction of late-time cosmology and null tests},
  {\emph{arXiv preprint arXiv:2111.11462} (2021) }.

\bibitem{gomez2021cosmological}
I.~G\'omez-Vargas, J.A.~V\'azquez, R.M.~Esquivel and R.~Garc\'\i{}a-Salcedo,
  \emph{{Cosmological Reconstructions with Artificial Neural Networks}},
  \href{https://arxiv.org/abs/2104.00595}{{\ttfamily 2104.00595}}.

\bibitem{wang2020reconstructing}
G.-J.~Wang, X.-J.~Ma, S.-Y.~Li and J.-Q.~Xia, \emph{Reconstructing functions
  and estimating parameters with artificial neural networks: A test with a
  hubble parameter and sne ia},
  \href{https://doi.org/https://doi.org/10.3847/1538-4365/ab620b}{\emph{Astrophysical
  Journal Supplement Series} {\bfseries 246} (2020) 13}.

\bibitem{escamilla2020deep}
C.~Escamilla-Rivera, M.A.C.~Quintero and S.~Capozziello, \emph{A deep learning
  approach to cosmological dark energy models},
  \href{https://doi.org/https://doi.org/10.1088/1475-7516/2020/03/008}{\emph{Journal
  of Cosmology and Astroparticle Physics} {\bfseries 2020} (2020) 008}.

\bibitem{graff2012bambi}
P.~Graff, F.~Feroz, M.P.~Hobson and A.~Lasenby, \emph{Bambi: blind accelerated
  multimodal bayesian inference}, {\emph{Monthly Notices of the Royal
  Astronomical Society} {\bfseries 421} (2012) 169}.

\bibitem{moss2020accelerated}
A.~Moss, \emph{Accelerated bayesian inference using deep learning},
  {\emph{Monthly Notices of the Royal Astronomical Society} {\bfseries 496}
  (2020) 328}.

\bibitem{hortua2020accelerating}
H.J.~Hortua, R.~Volpi, D.~Marinelli and L.~Malago, \emph{Accelerating mcmc
  algorithms through bayesian deep networks}, {\emph{arXiv preprint
  arXiv:2011.14276} (2020) }.

\bibitem{gomez2021neural}
I.~G{\'o}mez-Vargas, R.M.~Esquivel, R.~Garc{\'\i}a-Salcedo and
  J.A.~V{\'a}zquez, \emph{Neural network within a bayesian inference
  framework},
  \href{https://doi.org/https://doi.org/10.1088/1742-6596/1723/1/012022}{\emph{J.
  Phys. Conf. Ser.} {\bfseries 1723} (2021) 012022}.

\bibitem{mancini2021itcosmopower}
A.S.~Mancini, D.~Piras, J.~Alsing, B.~Joachimi and M.P.~Hobson,
  \emph{$\it{CosmoPower} \,$: emulating cosmological power spectra for
  accelerated bayesian inference from next-generation surveys},  2021.

\bibitem{baccigalupi2000neural}
C.~Baccigalupi, L.~Bedini, C.~Burigana, G.~De~Zotti, A.~Farusi, D.~Maino
  et~al., \emph{Neural networks and the separation of cosmic microwave
  background and astrophysical signals in sky maps}, {\emph{Monthly Notices of
  the Royal Astronomical Society} {\bfseries 318} (2000) 769}.

\bibitem{cmb_map}
M.A.~Petroff, G.E.~Addison, C.L.~Bennett and J.L.~Weiland, \emph{Full-sky
  cosmic microwave background foreground cleaning using machine learning},
  \href{https://doi.org/10.3847/1538-4357/abb9a7}{\emph{The Astrophysical
  Journal} {\bfseries 903} (2020) 104}.

\bibitem{clasificacion_2018}
J.~Pasquet-Itam and J.~Pasquet, \emph{Deep learning approach for classifying,
  detecting and predicting photometric redshifts of quasars in the sloan
  digital sky survey stripe 82},
  \href{https://doi.org/10.1051/0004-6361/201731106}{\emph{Astronomy \&
  Astrophysics} {\bfseries 611} (2018) A97}.

\bibitem{ribli2019improved}
D.~Ribli, B.{\'A}.~Pataki and I.~Csabai, \emph{An improved cosmological
  parameter inference scheme motivated by deep learning}, {\emph{Nature
  Astronomy} {\bfseries 3} (2019) 93}.

\bibitem{ishida2019machine}
E.E.~Ishida, \emph{Machine learning and the future of supernova cosmology},
  {\emph{Nature Astronomy} {\bfseries 3} (2019) 680}.

\bibitem{list2020galactic}
F.~List, N.L.~Rodd, G.F.~Lewis and I.~Bhat, \emph{Galactic center excess in a
  new light: Disentangling the $\gamma$-ray sky with bayesian graph
  convolutional neural networks}, {\emph{Physical Review Letters} {\bfseries
  125} (2020) 241102}.

\bibitem{dax2021real}
M.~Dax, S.R.~Green, J.~Gair, J.H.~Macke, A.~Buonanno and B.~Sch{\"o}lkopf,
  \emph{Real-time gravitational wave science with neural posterior estimation},
  {\emph{Physical review letters} {\bfseries 127} (2021) 241103}.

\bibitem{mcculloch1943logical}
W.S.~McCulloch and W.~Pitts, \emph{A logical calculus of the ideas immanent in
  nervous activity}, {\emph{The bulletin of mathematical biophysics} {\bfseries
  5} (1943) 115}.

\bibitem{rosenblatt1957perceptron}
F.~Rosenblatt and S.~Papert, \emph{The perceptron}, {\emph{A perceiving and
  recognizing automation, Cornell Aeronautical Laboratory Report} (1957) 85}.

\bibitem{minsky1969perceptron}
M.~Minsky and S.~Papert, \emph{Perceptron: an introduction to computational
  geometry}, {\emph{The MIT Press, Cambridge, expanded edition} {\bfseries 19}
  (1969) 2}.

\bibitem{rumelhart1986learning}
D.E.~Rumelhart, G.E.~Hinton and R.J.~Williams, \emph{Learning representations
  by back-propagating errors}, {\emph{nature} {\bfseries 323} (1986) 533}.

\bibitem{Ying_2019}
X.~Ying, \emph{An overview of overfitting and its solutions},
  \href{https://doi.org/10.1088/1742-6596/1168/2/022022}{\emph{Journal of
  Physics: Conference Series} {\bfseries 1168} (2019) 022022}.

\bibitem{avoid_overfit}
H.~Allamy, \emph{Methods to avoid over-fitting and under-fitting in supervised
  machine learning (comparative study)}, .

\bibitem{zhang2021dive}
A.~Zhang, Z.C.~Lipton, M.~Li and A.J.~Smola, \emph{Dive into deep learning},
  2021.

\bibitem{srivastava2014dropout}
N.~Srivastava, G.~Hinton, A.~Krizhevsky, I.~Sutskever and R.~Salakhutdinov,
  \emph{Dropout: a simple way to prevent neural networks from overfitting},
  {\emph{The journal of machine learning research} {\bfseries 15} (2014) 1929}.

\bibitem{louizos2017learning}
C.~Louizos, M.~Welling and D.P.~Kingma, \emph{Learning sparse neural networks
  through $ l\_0 $ regularization}, {\emph{arXiv preprint arXiv:1712.01312}
  (2017) }.

\bibitem{phaisangittisagul2016analysis}
E.~Phaisangittisagul, \emph{An analysis of the regularization between l2 and
  dropout in single hidden layer neural network},  in \emph{2016 7th
  International Conference on Intelligent Systems, Modelling and Simulation
  (ISMS)}, pp.~174--179, IEEE, 2016.

\bibitem{git}
``Full code repository.''
  \url{https://github.com/JuanDDiosRojas/Arts/tree/main/Deep%20Learning%20and%20its%20applications%20to%20cosmology}.

\bibitem{Escamilla:2021uoj}
L.A.~Escamilla and J.A.~Vazquez, \emph{{Model selection applied to
  non-parametric reconstructions of the Dark Energy}},
  \href{https://arxiv.org/abs/2111.10457}{{\ttfamily 2111.10457}}.

\bibitem{Keeley:2020aym}
R.E.~Keeley, A.~Shafieloo, G.-B.~Zhao, J.A.~Vazquez and H.~Koo,
  \emph{{Reconstructing the Universe: Testing the Mutual Consistency of the
  Pantheon and SDSS/eBOSS BAO Data Sets with Gaussian Processes}},
  \href{https://doi.org/10.3847/1538-3881/abdd2a}{\emph{Astron. J.} {\bfseries
  161} (2021) 151} [\href{https://arxiv.org/abs/2010.03234}{{\ttfamily
  2010.03234}}].

\bibitem{eqdiff}
I.~Lagaris, A.~Likas and D.~Fotiadis, \emph{Artificial neural networks for
  solving ordinary and partial differential equations},
  \href{https://doi.org/10.1109/72.712178}{\emph{IEEE Transactions on Neural
  Networks} {\bfseries 9} (1998) 987–1000}.

\bibitem{parciales1}
M.~Raissi, P.~Perdikaris and G.E.~Karniadakis, \emph{Physics informed deep
  learning (part i): Data-driven solutions of nonlinear partial differential
  equations},  2017.

\bibitem{parciales2}
M.~Raissi, P.~Perdikaris and G.E.~Karniadakis, \emph{Physics informed deep
  learning (part ii): Data-driven discovery of nonlinear partial differential
  equations},  2017.

\bibitem{DUFERA2021100058}
T.T.~Dufera, \emph{Deep neural network for system of ordinary differential
  equations: Vectorized algorithm and simulation},
  \href{https://doi.org/https://doi.org/10.1016/j.mlwa.2021.100058}{\emph{Machine
  Learning with Applications} {\bfseries 5} (2021) 100058}.

\bibitem{Padilla:2019mgi}
L.E.~Padilla, L.O.~Tellez, L.A.~Escamilla and J.A.~Vazquez, \emph{{Cosmological
  Parameter Inference with Bayesian Statistics}},
  \href{https://doi.org/10.3390/universe7070213}{\emph{Universe} {\bfseries 7}
  (2021) 213} [\href{https://arxiv.org/abs/1903.11127}{{\ttfamily
  1903.11127}}].

\bibitem{Vazquez:2020ani}
J.A.~V\'azquez, D.~Tamayo, A.A.~Sen and I.~Quiros, \emph{{Bayesian model
  selection on scalar $\epsilon$-field dark energy}},
  \href{https://doi.org/10.1103/PhysRevD.103.043506}{\emph{Phys. Rev. D}
  {\bfseries 103} (2021) 043506}
  [\href{https://arxiv.org/abs/2009.01904}{{\ttfamily 2009.01904}}].

\bibitem{Gonzalez:2008wa}
T.~Gonzalez, T.~Matos, I.~Quiros and A.~Vazquez-Gonzalez,
  \emph{{Self-interacting Scalar Field Trapped in a Randall-Sundrum Braneworld:
  The Dynamical Systems Perspective}},
  \href{https://doi.org/10.1016/j.physletb.2009.04.080}{\emph{Phys. Lett. B}
  {\bfseries 676} (2009) 161}
  [\href{https://arxiv.org/abs/0812.1734}{{\ttfamily 0812.1734}}].

\bibitem{hornik1989multilayer}
K.~Hornik, M.~Stinchcombe and H.~White, \emph{Multilayer feedforward networks
  are universal approximators}, {\emph{Neural networks} {\bfseries 2} (1989)
  359}.

\bibitem{GDconvergence}
R.M.~Gower, \emph{Convergence theorems for gradient descent}, {\emph{Lecture
  notes for Statistical Optimization} (2018) }.

\bibitem{nielsen2015neural}
M.A.~Nielsen, \emph{Neural networks and deep learning}, vol.~25, Determination
  press San Francisco, CA (2015).

\end{thebibliography}\endgroup



\providecommand{\href}[2]{#2}\begingroup\raggedright\endgroup

\appendix

\section{\textbf{Activation functions}} 
\label{app:deeplearning_fnact}

Artificial Neural Networks are known as universal approximators. The Approximation Theorem \cite{hornik1989multilayer} proves that, given a continuous function on a compact set of an $n$-dimensional space $ f: \mathds{R}^n \longrightarrow \mathds{R}^k$; then there exists a neural network with (at least) one hidden layer and a non-linear activation function, which approximates it with any desired degree of precision.

The activation function role is essential in Deep Learning; without them only linear transformations could be represented, no matter how many hidden layers and nodes are used. The choice of the activation functions depends on each case, or the problem being addressed, and the type of behaviour required in each layer of the network.

There are several activation functions that meet the features of being non-linear and continuous, however a few have been studied and applied more than others. Here are some common examples:

\begin{figure}[h!]
     \centering
     \makebox[8cm][c]{
     \begin{subfigure}{0.3\textwidth}
         \centering
         \includegraphics[width=\textwidth]{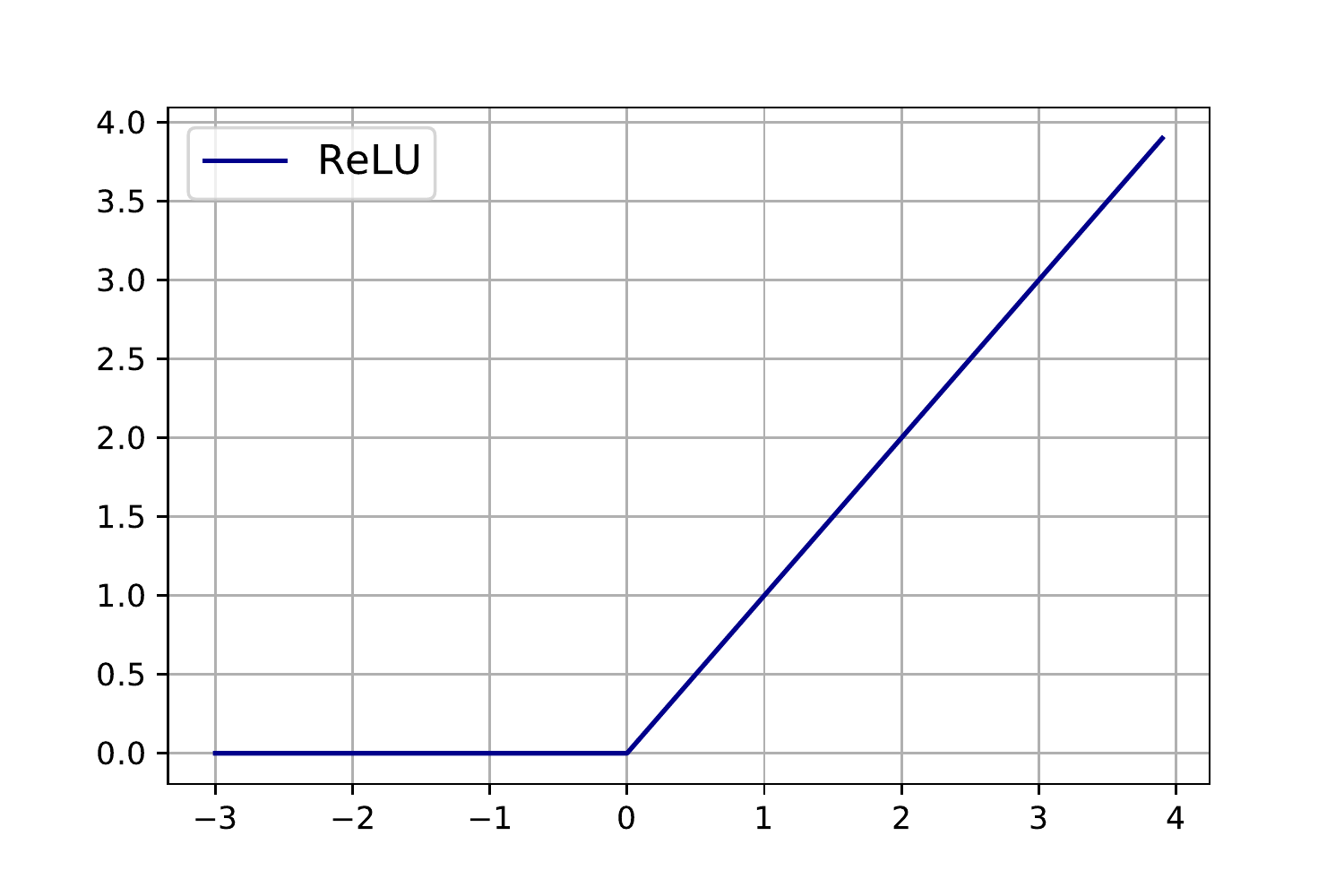}
         \caption{ReLu(x).}
         \label{fig:relu}
     \end{subfigure}
     \begin{subfigure}{0.3\textwidth}
         \centering
         \includegraphics[width=\textwidth]{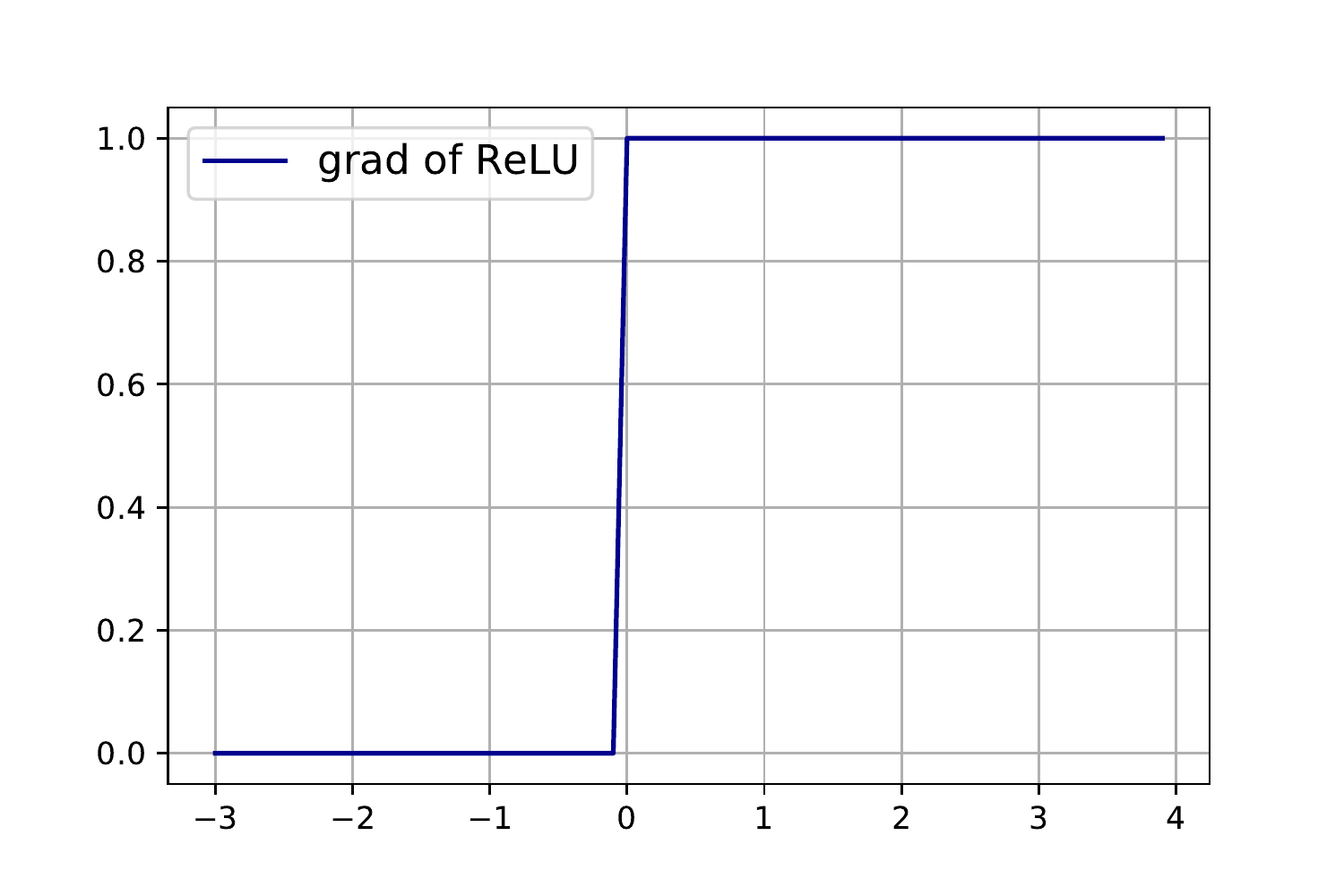}
         \caption{ReLu'(x).}
         \label{fig:drelu}
     \end{subfigure}
     }
     \makebox[8cm][c]{
     \begin{subfigure}{0.3\textwidth}
         \centering
         \includegraphics[width=\textwidth]{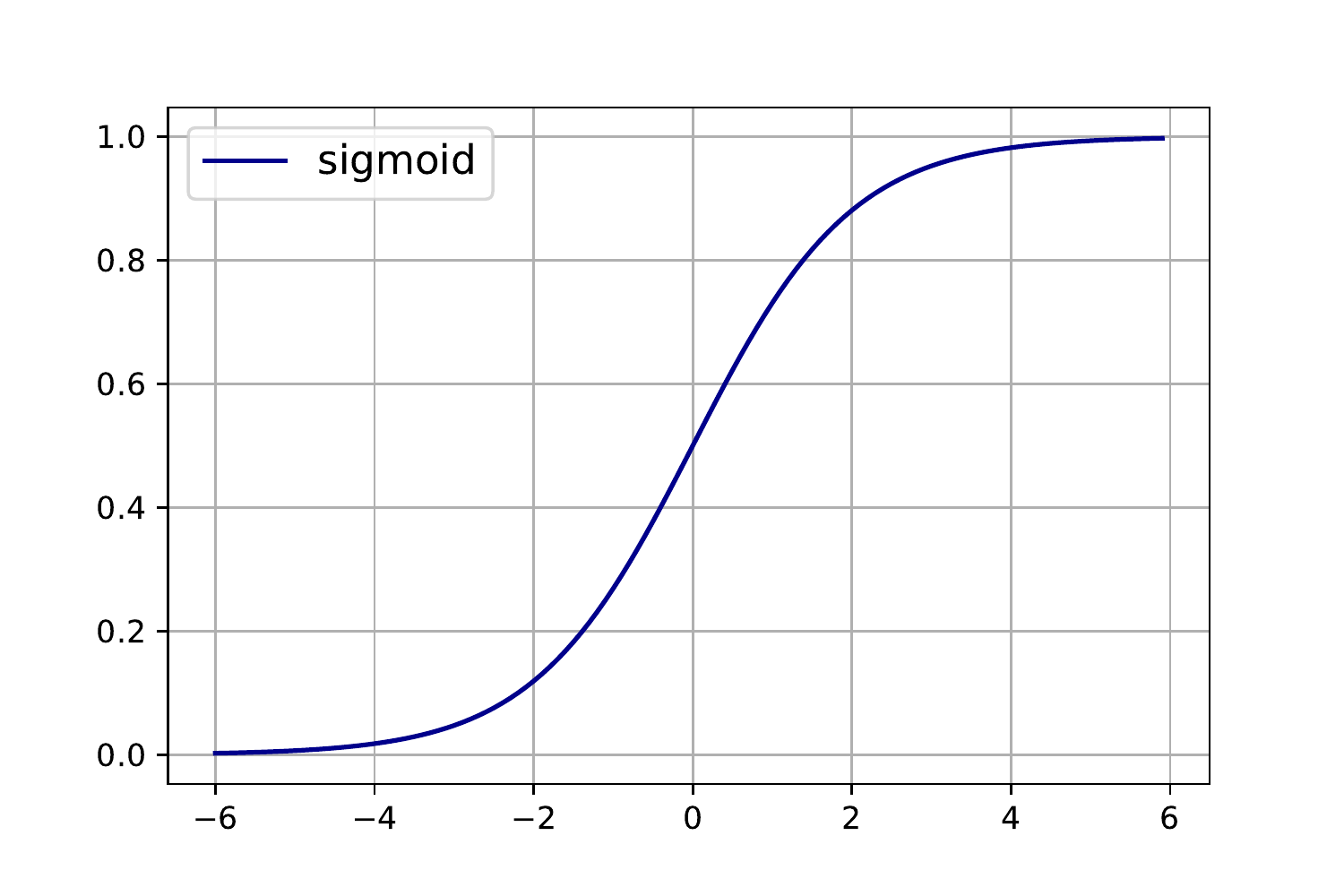}
         \caption{sigmoid(x).}
         \label{fig:grafsigmoid}
     \end{subfigure}
     \begin{subfigure}{0.3\textwidth}
         \centering
         \includegraphics[width=\textwidth]{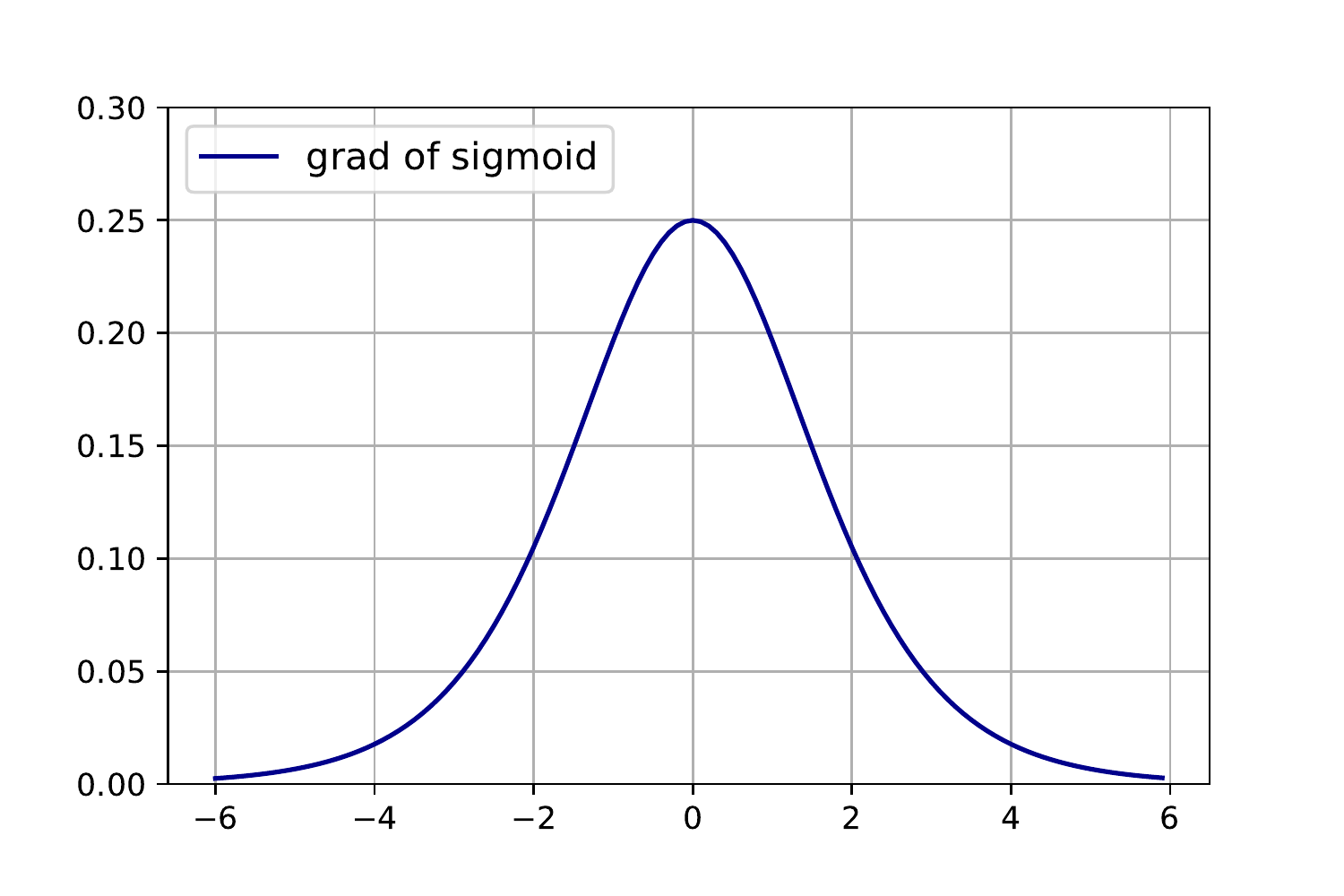}
         \caption{sigmoid'(x).}
         \label{fig:grafdsigmoid}
     \end{subfigure}
     }
     \makebox[8cm][c]{
     \begin{subfigure}{0.3\textwidth}
         \centering
         \includegraphics[width=\textwidth]{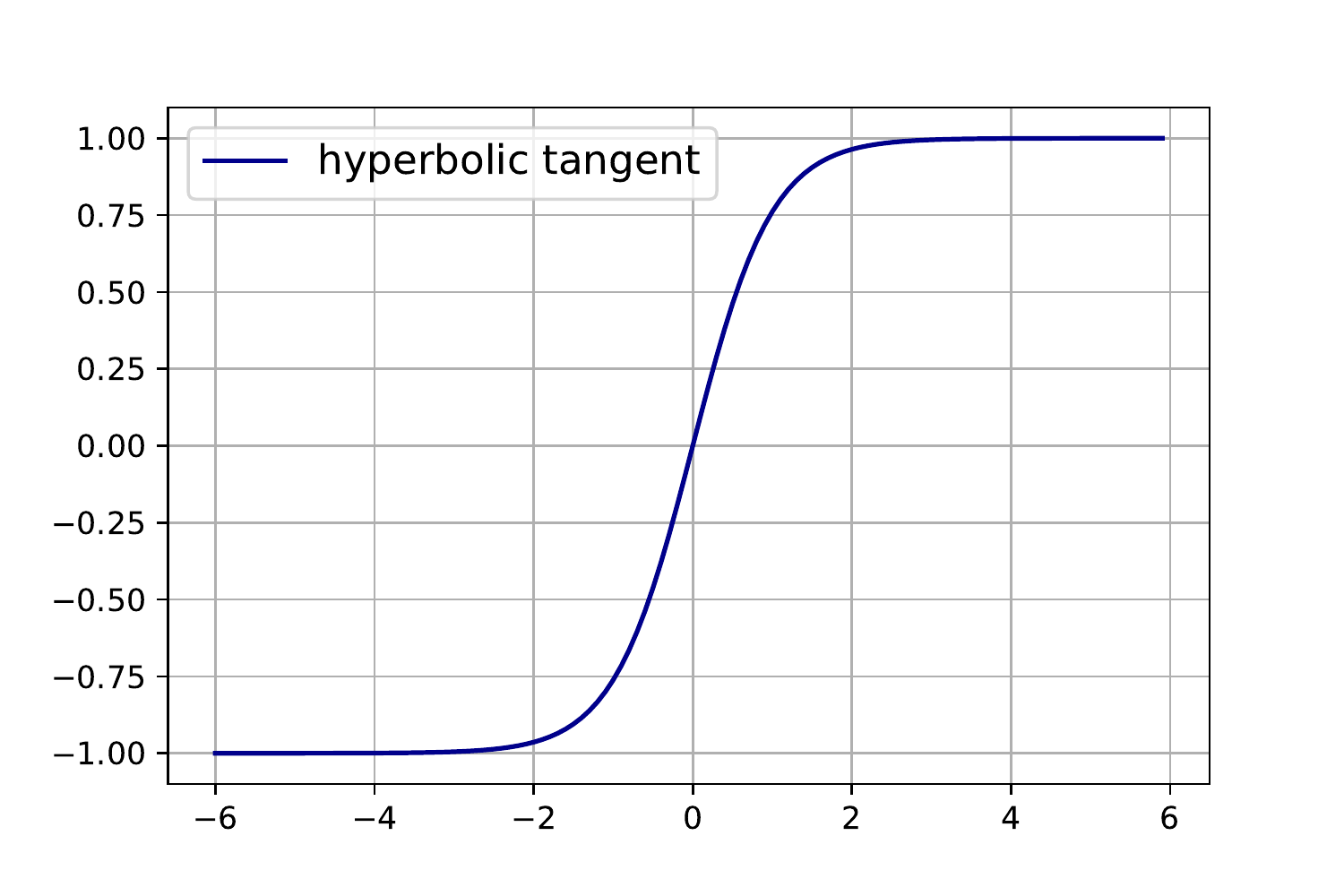}
         \caption{tanh(x).}
         \label{fig:graftanh}
     \end{subfigure}
     \begin{subfigure}{0.3\textwidth}
         \centering
         \includegraphics[width=\textwidth]{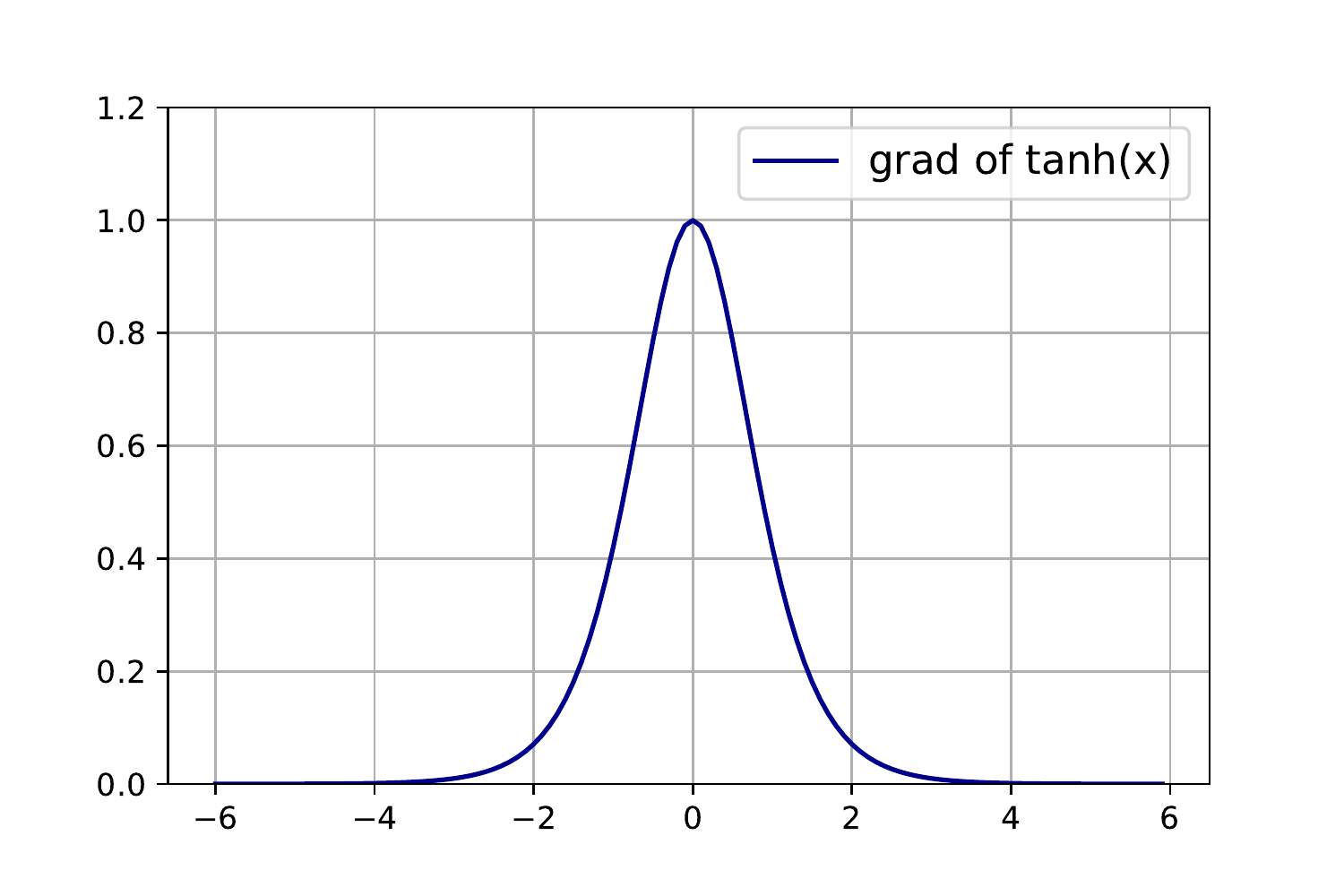}
         \caption{tanh'(x).}
         \label{fig:grafdtanh}
     \end{subfigure}
     }
    \caption{\footnotesize{Some common activation functions: Rectified Linear Unit (ReLu), sigmoid and hyperbolic tangent (tanh). All of them are nonlinear functions. Left: the curves of the activation functions, and right: their derivatives which are intensively used through backpropagation.}}
    \label{fig:three sin x}
\end{figure}

\begin{itemize}
    \item \textbf{ReLU}.- \textit{Rectified Linear Unit} (Fig.~\ref{fig:relu}) is the most popular and simple activation function. Given a number $x$, then  $$\text{ReLU}(x)=\text{max}\{0,x\}.$$
    Providing a very simple non-linear transformation over $\mathds{R}$. The ReLU function retains only positive elements and discards all negative by setting $x<0$ to 0. Although its derivative is undefined when $x=0$ (Fig.~\ref{fig:drelu}), it is not necessary to worry about it because the input values may never actually be all zero at the same time.
    $$\frac{d}{dx}\text{ReLU}(x)= \left\{
	       \begin{array}{ll}
		 0      & \mathrm{if\ } x < 0, \\
		 1 & \mathrm{if\ } x > 0. \\
	       \end{array}
	\right.$$
	
	Derivatives of the activation functions are a relevant part of the learning process. Therefore, it is important to consider their properties.
	
	\item \textbf{Sigmoid}.- The sigmoid function (Fig.~\ref{fig:grafsigmoid}) maps the real line to the interval $(0,1)$. The behaviour of this function was defined keeping in mind the behaviour of real neurons, which receive stimuli and communicate each other through pulses. The sigmoid function squashes the very negative $x$ to zero and if $x$ tends to infinity, its image will be mapped to $1$, that is, it is a good way to emulate a smoothed step function with 0 or 1. It is defined as:
    $$ \text{sigmoid} (x) = \frac{1}{1 + e^{- x}},$$
    and its derivative (Fig.~\ref{fig:grafdsigmoid}) reaches its maximum when $x=0$, and when it moves away from this value, it tends to zero
    $$ \frac{d}{dx}\text{sigmoid} (x) = \frac{e^{- x}}{(1 + e ^{-x})^2}=$$
    $$\text{sigmoid}(x)(1-\text{sigmoid} (x)).$$
    The second definition is better in computational terms.
    
    \item \textbf{Hyperbolic tangent}.- This function has a similar behaviour to the sigmoid function, but it provides negative values. In the same way, it maps the set of real numbers to the interval $(-1,1)$. For the points close to $ 0 $, the hyperbolic tangent function ($ \text{tanh} $) has an almost linear behaviour and it is symmetric with respect to the Y axis (Figure \ref{fig:graftanh})
    $$\text{tanh}(x)=\frac{1-e^{-2x}}{1+e^{-2x}}.$$
    Regarding its derivative (Fig.~\ref{fig:grafdtanh}), the behaviour is similar to the derivative of the sigmoid
    $$\frac{d}{dx}\text{tanh}(x)=1-\text{tanh}(x)^2.$$
\end{itemize}

\section{Gradient descent}
\label{app:perceptron_descenso}

Even though there exists a broad diversity of optimisation algorithms used to minimise the cost function, in Deep Learning the gradient descent is the most popular. 
It is versatile and may be generalised to multivariable functions. 
For example, given a scalar differentiable function $f: \: \mathds{R}^n\longrightarrow \mathds {R}$, the vector whose components are the partial derivatives of $f$, is called \textbf{Gradient} $\nabla f(w)$. In addition, the gradient has some interesting geometric properties:
\begin{itemize}
    \item  $\nabla f(w_0)$ is orthogonal to the level curve $f(w)=k$ at the point $w_0$.
    \item $\nabla f(w)$ points to the direction of the maximum increase of $f$.
    \item Conversely, $-\nabla f(w)$ points to the direction of the maximum decrease.
\end{itemize}
The gradient descent algorithm relies on the third property. Let $f$ be a function such that it has a minimum at $x_0$, with a point $x$ located at certain region of the domain. 
By taking a step in the $-\nabla f(x)$ direction, the new position is one step closer to the critical point $x_0$, and with sufficient number of iterations, the value for which $f$ is minimised can be found.
The following equation synthesises the mechanism of the gradient descent:
\begin{equation}
     w '\longrightarrow w-\nabla f(w),
     \label{eq:gradiente}
\end{equation}
where $x'$ is the new point to be taken by the algorithm and $x$ is the current point. It can be noticed that one of the relevant advantages of the gradient descent, over other types of optimisation algorithms, is that it does not use second derivatives, and this property makes it more suitable. 

An important parameter within the gradient descent is the \textit{learning rate}, commonly denoted by $\eta$. This parameter determines the step size that it takes in each iteration, therefore this value is very important because it influences whether or not the algorithm is able to converge to the target value in an appropriate way. The learning rate is usually a positive real constant number ($\eta \in [0,1]$) by which the gradient is multiplied. When $\eta$ is incorporated into Eq.~(\ref{eq:gradiente}), this results in:
\begin{equation}
    w'\longrightarrow w- \eta \nabla f(w).
    \label{eq:gradrate}
\end{equation}

\begin{figure}[t!]
     \centering
     \begin{subfigure}{0.4\textwidth}
         \centering
         \includegraphics[trim = 10mm  0mm 0mm 0mm, clip, width=8.cm, height=4.5cm]{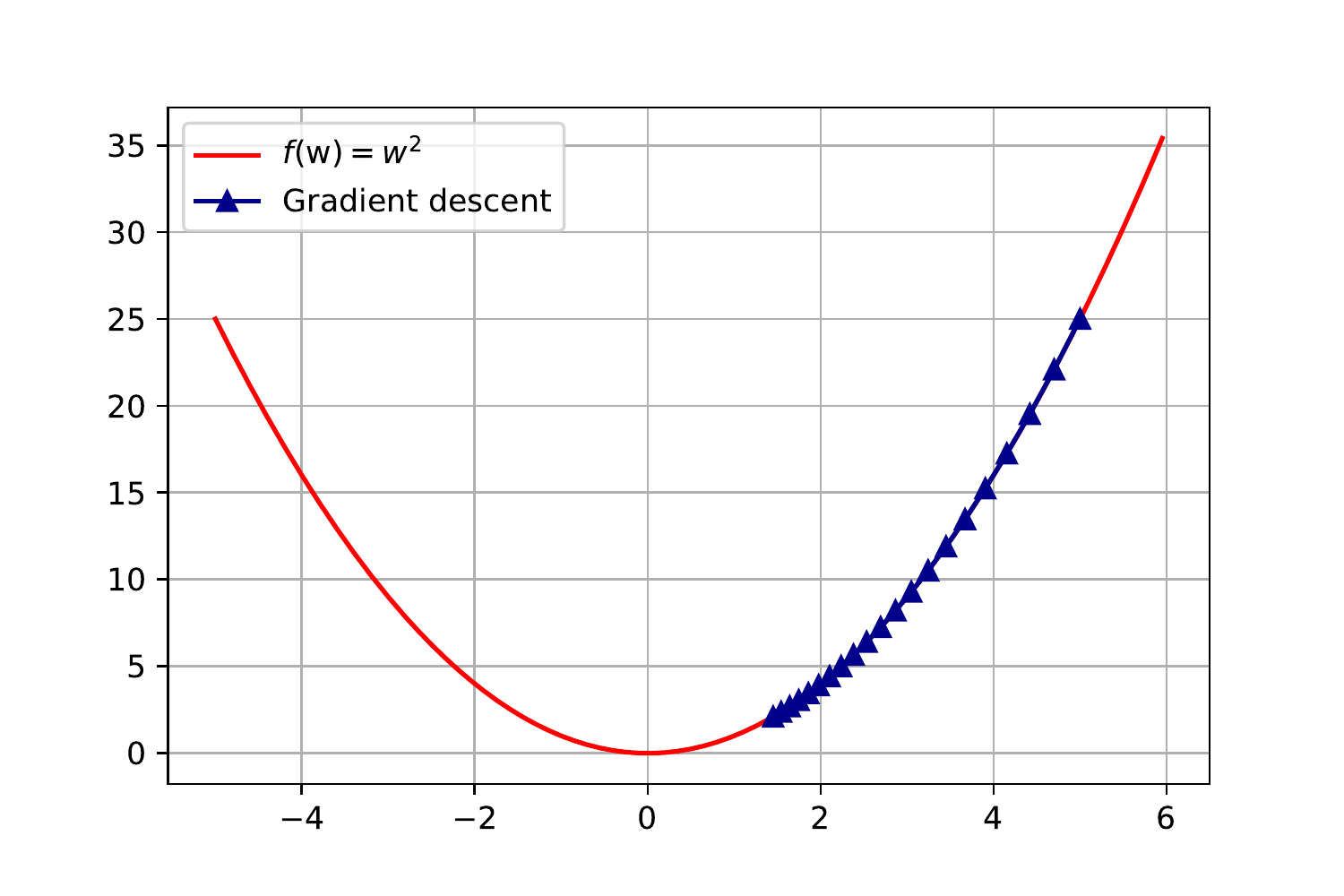}
         \caption{Too small $\eta$.}
         \label{fig:paso pequeño}
     \end{subfigure}
     \begin{subfigure}{0.4\textwidth}
         \centering
         \includegraphics[trim = 10mm  0mm 0mm 0mm, clip, width=8.cm, height=4.5cm]{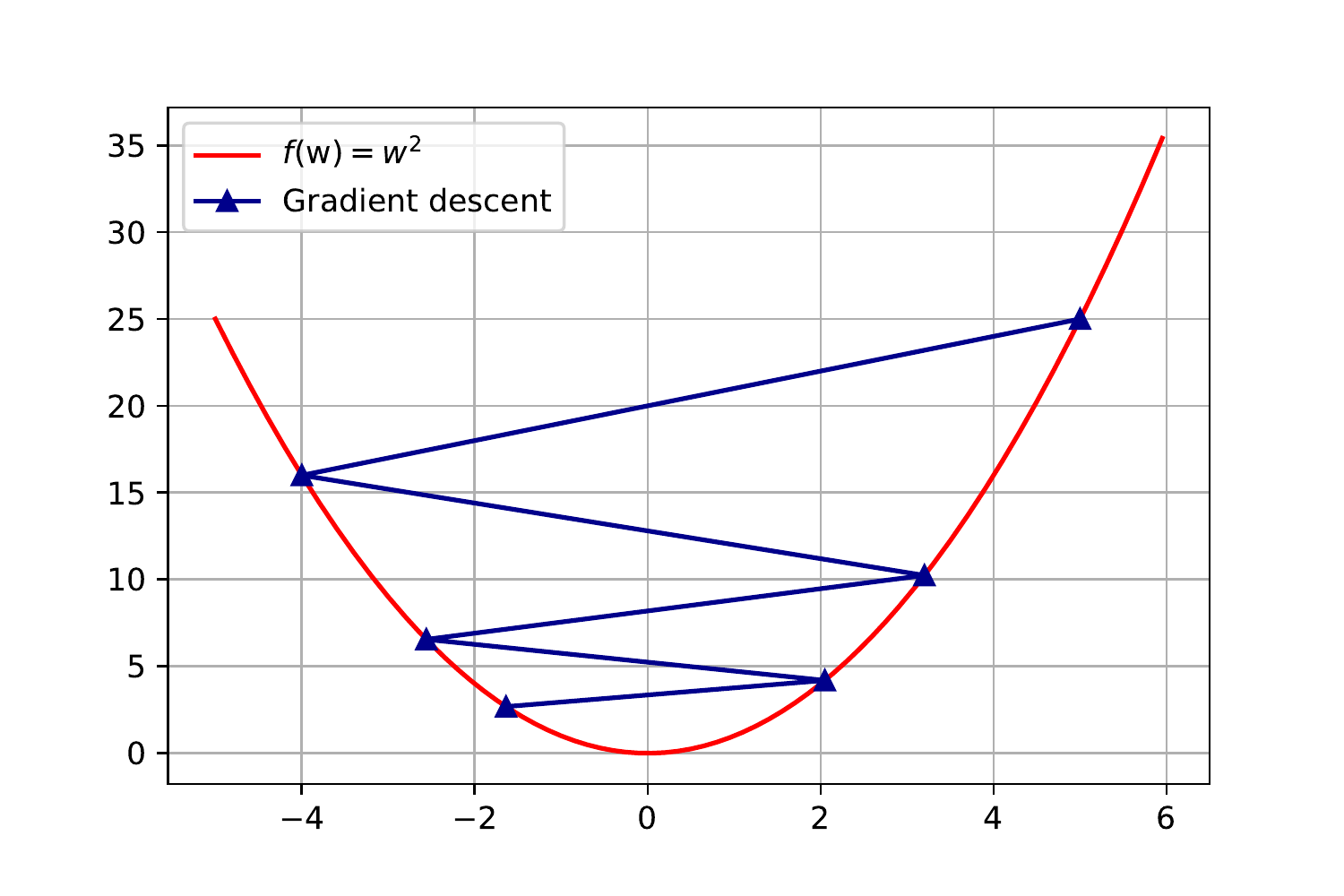}
         \caption{Too large $\eta$.}
         \label{fig:paso grande}
     \end{subfigure}
     \begin{subfigure}{0.4\textwidth}
         \centering
         \includegraphics[trim = 10mm  0mm 0mm 0mm, clip, width=8.cm, height=4.5cm]{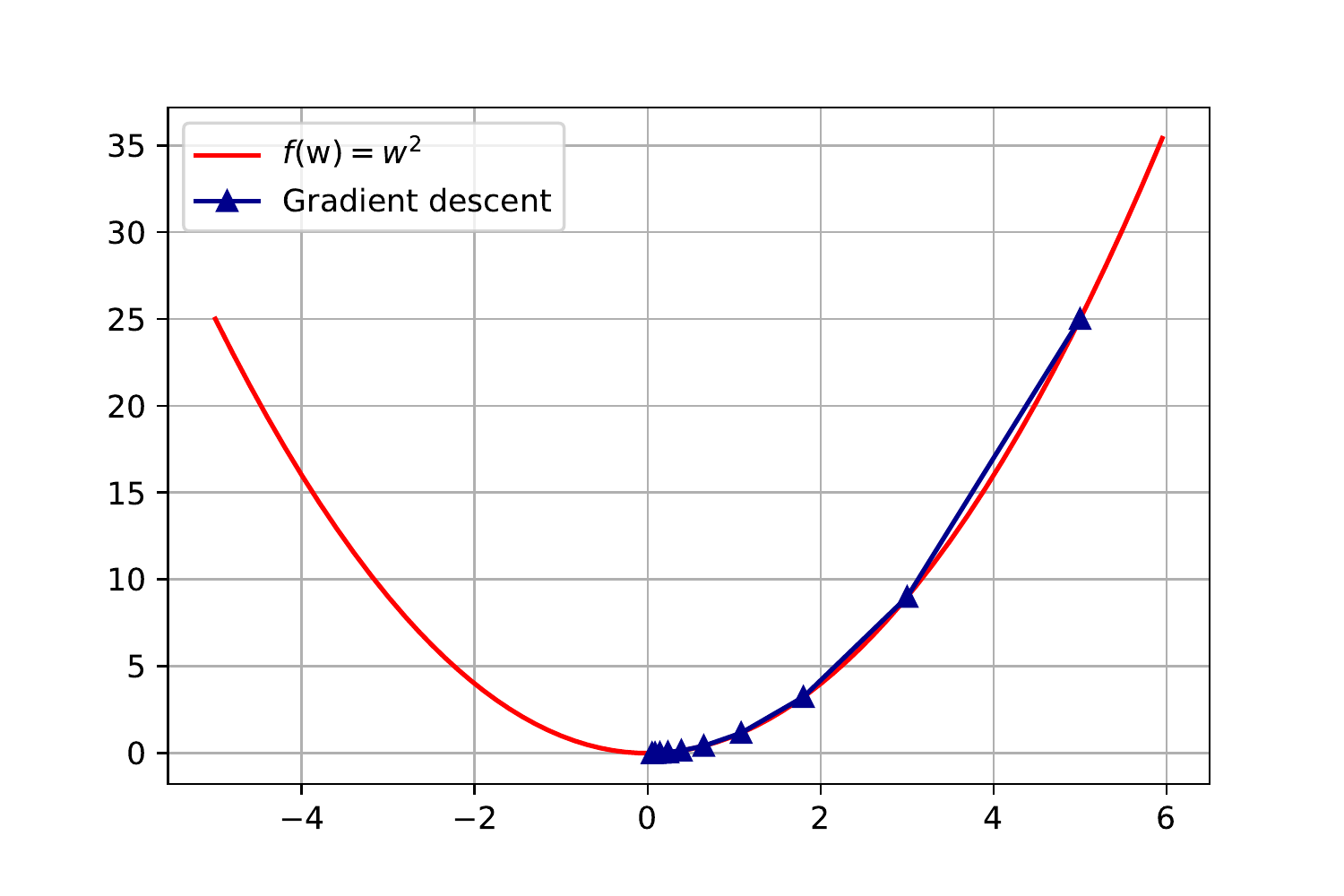}
         \caption{Optimal $\eta$.}
         \label{fig: paso adecuado}
     \end{subfigure}
    \caption{\footnotesize{Three different selections of the learning rate and their impact on the gradient descent when applying to minimise a function.}}
    \label{fig:learning rate}
\end{figure}

What should be the right value for $\eta$? It depends on the function to be minimised, different types of problems correspond to different values. There are some cases where it is more useful to consider a dynamic learning rate, whose size increases and shrinks as approach the minimum of the function. 
In this case, the step size in each iteration depends on two factors: the norm of the gradient vector and the $\eta$ value. As we get closer to the critical point, the step size tends to get shorter; even though $\eta $ is constant, since at the critical point the gradient tends to have a zero vector. In addition, there are three cases to consider for the learning rate magnitude:
\begin{itemize}
    \item \textbf{Too small $\eta$}: If $\eta$ is very small, the step size will be too short, then it might increase the total computation time to a very large extent or could even fail to converge to the desired point (see Fig.~\ref{fig:paso pequeño}).
    
    \item \textbf{Too large $\eta$}: A very large learning rate can cause an exploding gradient and diverge from the minimum, or the algorithm may bypass the local minimum and overshoot (see Fig.~\ref{fig:paso grande}).
    
    \item \textbf{Optimal $\eta$}: A proper learning rate ensures that the algorithm can converge to the minimum of the function on a reasonable number of attempts, which also reduces the computation time of the process (Fig.~\ref{fig: paso adecuado}).
\end{itemize}

We can implement gradient descent by following these three steps:
\begin{enumerate} 
\item Compute partial derivatives of each variable $x_k$ and evaluate them at a random starting point $w_0$ in the domain of the function, $\frac{\partial f}{\partial w_k} (w_0) $.

\item The gradient of the function must be constructed. It is recommended to define a maximum value for the norm of the gradient, but preserving their direction, to avoid exploding when doing the iterations. 

\item  Apply Eq.~(\ref{eq:gradrate}) to the initial point $ x_0 $ and the process is repeated for the new point until it is closed enough to the minimum.
\end{enumerate}

\begin{figure} [t!]
    \centering
    \includegraphics [trim = 0mm  0mm 0mm 0mm, clip, width=8.5cm, height=4.5cm]{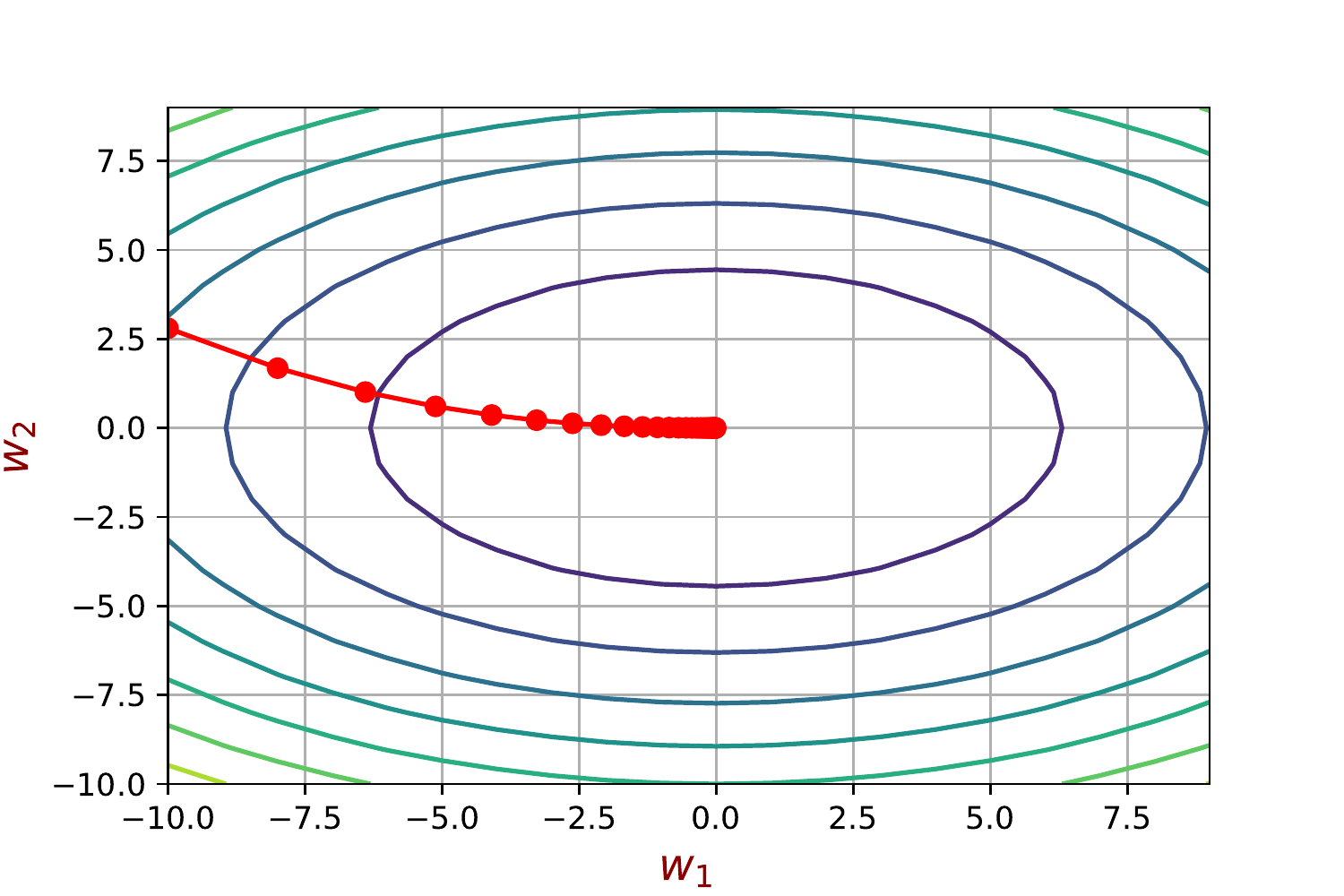}
    \caption {\footnotesize{Red line displays the path taken by the gradient descent in the $f$ domain to find its minimum. The ellipses correspond to the level curves of $f$.}}
    \label{fig:camino_gradiente}
\end{figure}

 In order to summarise the ideas, we apply the gradient descent algorithm to the function $f(w_1,w_2) = w_1^2 + w_2^2$ to find its minimum; using a learning rate $\eta = 0.1$ and a starting point $w_0 = (w_{1,0}, w_{2,0}) = (-10, 2.8) $.
 In Fig.~\ref{fig:camino_gradiente} we plot the 2D trajectory the algorithm follows during the process. It can be seen the steps on the level curves of $f$ until reaching the point where the  minimum is located, in this case $ (w_1,w_2)=(0,0) $.

In Fig.~\ref{fig: paso adecuado}, it can be noticed that the gradient descent found its minimum easily, although this is not always possible for every function. In general, there are difficulties with functions that have local minima or are not convex.
The convergence of the algorithm can be guaranteed if $ f $ meets a couple of conditions \cite{GDconvergence}:
\begin {enumerate}
    \item It is convex.
    \item  Its gradient vector is Lipschitz continuous, that is, for  a real positive number $L$: \\
    $ \norm {\nabla f (w) - \nabla f (v)} <L \norm {w-v} $.
\end{enumerate}

The gradient descent can be applied to many cost functions satisfying these conditions, which ensures that the desired critical point can be obtained. An example of a function that satisfies the previous characteristics is the Mean Squared Error (MSE). Therefore, we can use the MSE on the perceptron and then minimise it with the help of the gradient descent. 

In practice, there are variants of the gradient descent algorithm that are more efficient in some scenarios, for example, Stochastic Gradient Descent and Batch Gradient Descent. 

\section{Backpropagation equations}
\label{app:backpropagation}

Considering the mean squared error as cost function, which compares a prediction of the neural network $a^L$ with the expected value $Y$, we have the following:
\begin{equation}
    C=\frac{1}{2m} \norm{Y-a^L}^2=\frac{1}{2m}\sum_j \left(Y_j-a^L_j\right)^2,
\label{eq:error componentes}
\end{equation}
where $m$ is the number of samples. In order to optimise the Eq.~(\ref{eq:error componentes}), the Stochastic Gradient Descent algorithm can be applied iteratively until the minimum is reached. The weights $w_{ij}^k$ and biases $b_j^l$ must be updated in each iteration to improve the predictions of the MLP and to reduce the value of the cost function. 
However, considering a problem with multiple variables influencing the cost function, from Eq.~(\ref{eq:gradrate}) we have the new weights values:
\begin{equation}
    W^l \longrightarrow W^l- \eta \:\partial_{W^l} \left( \frac{1}{2}\sum_j \left(Y_j-a^L_j\right)^2 \right),
    \label{eq:grad pesos}
\end{equation}
 and the following bias values:
\begin{equation}
     b^l \longrightarrow b^l- \eta \:\partial_{b^l} \left(\frac{1}{2}\sum_j \left(Y_j-a^L_j\right)^2\right).
     \label{eq:grad bias}
\end{equation}
%
With Eqs.~(\ref{eq:grad pesos}) and (\ref{eq:grad bias}) it is possible to know how to update the MLP hyperparameters. However, these expressions only depend on the output layer parameters and the information of intermediate layers is unclear.
Therefore, it is useful to know how much the weights and biases influence the cost function. The algorithm that allows to propagate the error of the output layer to the elements of the previous layers is known as the \textit{backpropagation}. Thus, it is necessary to use the chain rule to find the partial derivatives in the Eqs.~(\ref{eq:grad pesos}) and (\ref{eq:grad bias}). By doing this, the \textit{fundamental equations of backpropagation} are obtained, which is a key concept in Deep Learning (see more details in \cite{nielsen2015neural}).

The backpropagation equations are derived as follows. First of all, it is necessary to consider the change in the cost function with respect to the weighted sum of the last layer, using the chain rule we have:
$$\frac{\partial C}{\partial z_j^L}=\frac{\partial C}{\partial a_j^L} \frac{\partial a_j^L}{\partial z_j^L},$$
$$
\qquad
\frac{\partial C}{\partial a_j^L}= \left( a_j^L-Y_j \right),
\qquad
 \frac{\partial a_j^L}{\partial z_j^L}=\sigma '(z^L_j).$$
Thus, considering these components we can obtain $\frac{\partial C} {\partial z^L}$, which is equivalent to:
\begin{equation}
    \frac{\partial C}{\partial z^L}=  \left( a^L-Y \right) \odot \sigma '(z^L)=\delta^L,
\label{eq:back 1}
\end{equation}
where $\odot$ represents the product component by component between two vectors of equal size. The expression in equation (\ref{eq:back 1}) denoted by $\delta^L$, often called \textit{imputed error}, can be calculated for any cost function applied to the neural network (MSE in this case).
The next step is to analyse how the cost function changes with respect to the variations of the parameters $W^L$, $b^L$ in the last layer:

$$\frac{\partial C}{\partial W^L} = \underbrace{\frac{\partial C}{\partial a^L} \frac{\partial a^L}{\partial z^L}}_{\delta^L} \frac{\partial z^L}{\partial W^L}, 
\qquad  
\frac{\partial C}{\partial b^L} = \underbrace{\frac{\partial C}{\partial a^L} \frac{\partial a^L}{\partial z^L}}_{\delta^L} \frac{\partial z^L}{\partial b^L}.$$
Note that in both equations, the first two factors are nothing other than the error $\delta^L$, furthermore:
$$\frac{\partial z^L}{\partial W^L} = {a^{L-1}}^T,
\qquad
\frac{\partial z^L}{\partial b^L} =1.$$
Therefore, substituting these values, the partial derivatives are as followed:
$$\frac{\partial C}{\partial W^L} = {a^{L-1}}^T \delta^L,
\qquad
\frac{\partial C}{\partial b^L} =\delta^L.$$

These last two equations are the partial derivatives that the gradient descent needs to update $W^L$ and $b^L$, but this is only for the last layer, therefore it is necessary to find the change of $C$ due to the previous layers. This is relatively straightforward, since it is enough to calculate the change with respect to the $L-1$ layer to find the one of the previous layers. For $ L-1 $:
$$\frac{\partial C}{\partial W^{L-1}} = 
\underbrace{\frac{\partial C}{\partial a^L} 
\frac{\partial a^L}{\partial z^L}}_{\delta^L} 
\underbrace{\frac{\partial z^L}{\partial a^{L-1}}}_{{W^L}^T} 
\underbrace{\frac{\partial a^{L-1}}{\partial z^{L-1}}}_{\sigma'(z^{L-1})}
\underbrace{\frac{\partial z^{L-1}}{\partial W^{L-1}}}_{{a^{L-2}}^T},$$
$$\frac{\partial C}{\partial b^{L-1}} = 
\underbrace{\frac{\partial C}{\partial a^L} 
\frac{\partial a^L}{\partial z^L}}_{\delta^L} 
\underbrace{\frac{\partial z^L}{\partial a^{L-1}}}_{{W^L}^T} 
\underbrace{\frac{\partial a^{L-1}}{\partial z^{L-1}}}_{\sigma'(z^{L-1})}
\underbrace{\frac{\partial z^{L-1}}{\partial b^{L-1}}}_{1}.$$
Notice that the first four factors of both equations correspond to the derivative
$$\delta^{L-1}=\frac{\partial C}{\partial z^{L-1}}=\frac{\partial C}{\partial a^L} \frac{\partial a^L}{\partial z^L}\frac{\partial z^L}{\partial a^{L-1}} \frac{\partial a^{L-1}}{\partial z^{L-1}},$$
which is equivalent to the imputed error of the $L-1$ layer. However, these partial derivatives have already been deduced and their values are known, so the imputed error is:
$$\delta^{L-1}=\delta^L {W^L}^T \odot \sigma'\left({z^{L-1}}\right).$$

Even though this expression was derived for the imputed error of layer $L-1$, it can be noticed that, if we were going to find the error corresponding to the layer $L-2$, we would obtain the same relation between this one and the layer that precedes it, thus this relation can be generalised for any layer $l-1$. Moreover, doing a change of indices, we obtain the expression:
\begin{equation}
    \delta^l=\delta^{l+1}{W^{l+1}}^T\odot\sigma'\left(z^l\right).
\label{eq:back 2}
\end{equation}
This same generalisation and index arrangement is valid for the expressions  $\frac{\partial C}{\partial W^{L-1}}$ and $\frac{\partial C}{\partial b^{L-1}}$, from which the expressions for the derivatives of the hidden layers are obtained:
\begin{equation}
     \frac{\partial C}{\partial W^l}={a^{l-1}}^T\delta^l,
\label{eq:back 3}
\end{equation}
\begin{equation}
    \frac{\partial C}{\partial b^l}=\delta^l.
\label{eq:back 4}
\end{equation}

The Eqs.~(\ref{eq:back 1}) - (\ref{eq:back 4}) are known as the \textit{fundamental backpropagation equations}, since they can update the parameters in each layer by applying the gradient descent mentioned in Eqs.(\ref{eq:grad pesos}) and (\ref{eq:grad bias}). Furthermore, if the Eqs.~(\ref{eq:back 3}) and (\ref{eq:back 4}) are substituted, we obtain:
\begin{equation}
    W^l \longrightarrow W^l-\eta \nabla_{W^l}C = \eta\sum_x {a^{l-1}_x}^T\delta^l_x,
\label{eq:gradiente pesos}
\end{equation}

\begin{equation}
     b^l \longrightarrow b^l-\eta \nabla_{b^l}C = \eta \sum_x\delta^l_x,
\label{eq:gradiente bias}
\end{equation}
where the subscript $x$ refers to the examples contained in the training set $X$, with $x \in X$.

The backpropagation is a crucial process for learning in a neural network as the parameters are updated following this algorithm. To implement the backpropagation in a neural network, the following steps are necessary:
\begin{enumerate}
\item Generate random values for $W^l $ and $b^l$. Then compute the corresponding output $a^L$ by forward propagation.
\item Compute the value of the cost function  (\ref{eq:error componentes}), then obtain the imputed error $\delta^L$ with respect to the output of the neural network (Eq.~\ref{eq:back 1}).
\item Compute the imputed error for the previous layers $ L-1, L-2 ... $ using the Eq.~(\ref{eq:back 2}).
\item Use the Eqs.~(\ref{eq:gradiente pesos}) and (\ref{eq:gradiente bias}) to update the MLP parameters.
\item Once the new parameters are in place, iterate this procedure until the cost function reaches a very small value.
\end{enumerate}
\end{document}